
\mag=\magstep1
\hsize 6.5truein
\vsize 8.9truein
\lefthyphenmin=3
\clubpenalty 10000
\brokenpenalty 10000
\displaywidowpenalty 10000
\pretolerance 1000

\def\etal{{\it et~al.}}
\def\ul#1{$\underline{\smash{\vphantom{y}\hbox{#1}}}$}
\def\lap{\hbox{~{\lower -2.5pt\hbox{$<$}}\hskip -8pt\raise
-2.5pt\hbox{$\sim$}}}
\def\gap{\hbox{~{\lower -2.5pt\hbox{$>$}}\hskip -8pt\raise
-2.5pt\hbox{$\sim$}}}
\def\degr{^\circ}

\baselineskip 20pt

\centerline{\bf COSMOLOGICAL EXPERIMENTS IN CONDENSED}
\centerline{\bf MATTER SYSTEMS}
\medskip
\centerline{\bf by}
\centerline{\bf W. H. Zurek}
\centerline{Theoretical Astrophysics}
\centerline{T-6, MS B288}
\centerline{Los Alamos National Laboratory}
\centerline{Los Alamos, NM  87545}
\baselineskip 15pt
\vfill\eject
\centerline{\bf ABSTRACT}
\medskip

\nobreak
{\narrower\smallskip
\bf Topological defects are thought to be left behind by the
cosmological phase transitions which occur as the universe expands and
cools.   Similar processes can be studied in the phase transitions which
take place in the laboratory:  ``Cosmological'' experiments in superfluid
He$^4$ and in liquid crystals were carried out within the past few years,
and their results shed a new light on the dynamics of the
defect-formation process.  The aim of this paper is to review the key
ideas behind this cosmology - condensed matter connection and to
propose new experiments which could probe heretofore unaddressed
aspects of the topological defects formation process.
\par}
\medskip
\centerline{\ul{\hskip 3truein}}
\bigskip
\baselineskip 20pt
\noindent
{\bf 1.\quad  INTRODUCTION}

\nobreak
``Big Bang'' leaves a radiation-dominated universe at a temperature
close to the Planck temperature.  As the resulting primordial fireball
expands, temperature falls, thus precipitating a sequence of
symmetry-breaking phase transitions, the last of which converts
quark-gulon plasma into nucleons.  In the intervening period of
expansion, many such transitions could have taken place.  The exact
number as well as the nature of these transitions is not known, but it is
often suggested that the origin of the observed structures---galaxies,
clusters, large-scale voids---is intimately related to the phase
transitions which may have taken place very early in the history of the
universe.  For example, inflationary cosmology (Linde, 1990) traces the
origins of the universe ``as we know it'' to the era in which the very
first symmetry breaking in the GUT scheme was taking place.  Indeed, in
many models inflation happens {\it in the course\/} of the (usually
second-order) phase transition (Guth, 1981; Albrecht and Steinhardt,
1982; Linde, 1982), leaving behind density perturbations which can
act as seeds of the structures seen today in the universe.

In addition to inflation there may have been many other phase
transitions, which could have left their imprints on the mass
distribution of the universe.  In particular, for a large class of field
theories the state left after the phase transition contains topological
defects---fragments of the original high-energy pre-transition
(symmetric) ``false vacuum'' which became locked in within the
topologically stable configurations of the new, broken symmetry ``true
vacuum'' (Zeldovich, Kobzarev, and Okun, 1974;  Kibble, 1976).    Such
frozen-out relics of the distant past come in three basic varieties:
monopoles, (cosmic) strings, and domain walls.  The kind of defects
which will be left behind by the transition depends on the interplay
between the symmetry which is being broken during the phase
transition (more specifically, on the homotopy group of the manifold of
the degenerate broken symmetry vacua) and the symmetries of the
space in which the evolution is taking place (Kibble, 1976).

Monopoles are an inevitable outcome of the GUT-era phase
transition and a major threat to cosmological models.  Their mass
would dominate the universe and force it to recollapse well before the
present era.  Inflationary cosmology was in part motivated by the
desire to dilute monopoles (Guth, 1981).  Cosmic strings are ``optional''
and may result in density perturbations not too different from those
produced by inflation (Vilenkin, 1985).  Domain walls are again a
threat---they are simply too massive to be consistent with either the
observed structures or the (relatively small) perturbations of the
cosmic microwave background.

The process of formation of topological defects---the main subject of
this paper---is perhaps the least understood (and most interesting)
aspect of their evolution.  The nature of this process depends on the
nature of the phase transition.  In the course of the first-order phase
transitions, a new broken symmetry phase usually {\it nucleates\/}
after the temperature falls some distance below the critical
temperature $T_c$.  Separated islands of the new phase form
independently and expand, resulting in a local selection of the broken
symmetry vacuum.  As a result of these independent selections, the
final configuration may not be able to get rid of the locked-out
fragments of the original, pre-transition vacuum.  First-order phase
transitions can also occur through the process known as spinodal
decomposition which by-passes the post-nucleation epoch of
coexistence of the two phases.

Second-order phase transitions are still more interesting.  There the
phase transformation occurs (approximately) simultaneously
throughout the volume.  However, unless the critical temperature is
traversed infinitesimally slowly, the resulting broken symmetry phase
will contain many distinct regions with different choices of the local
vacuum.  This is because the selection of the new vacuum has to be
somehow communicated if the same choice is to be made elsewhere,
and the speed at which this information can be propagated is finite.
Hence, appearance of the topological defects is---as was observed by
Kibble in his seminal paper (Kibble, 1976)---a direct consequence of
causality.

In the cosmological setting the ultimate causal limit on the size of
the approximately uniform patches of the new vacuum is set by the
size of the horizon at the time when the transition is taking place
(Kibble, 1976).  In the laboratory this ultimate limit is not relevant.
Indeed, also in the cosmological context the process which will set the
size of the uniform patches of the broken symmetry vacuum is likely to
be associated with the dynamics of the order parameter rather than
with the causal horizon (which can supply but an ultimate upper limit).
In the original paper, Kibble appealed to the process of thermal
activation, which is thought to occur in the course of the second-order
phase transitions below the critical temperature,
$T_c$, but above the so-called Ginzburg temperature $T_G$.  When the
temperature $T$ of the system falls in between $T_G$ and $T_c$, free
energy required to ``flip'' locally uniform (correlation length-sized)
volume of the new phase is less than the available thermal energy per
degree of freedom,
$\sim k_B T$.  Hence, while locally the new vacuum has been selected,
this choice is not really permanent---transitions between different
possible vacua occur on a relaxation time scale, which can be much
shorter than the time over which temperature traverses the region
between $T_c$ and $T_G$.  Consequently, it was originally postulated
that the initial density of topological effects is set by the correlation
length at the Ginzburg temperature, $\xi (T_G)$ (Kibble, 1976).

In the proposal for the early universe-like experiments
suggested as a superfluid test of the cosmological scenario (Zurek,
1984; 1985; 1993) a different alternative was put forward.   
It focuses on the nonequilibrium aspect of the phase transition and 
predicts that the characteristic domain size should be set by the
critical slowing down.  Order parameter can adjust only on a relaxation
time scale which diverges at $T = T_c$.  Thus, as the critical
temperature is approached from above, at a certain instant $\hat t$
evolution of the perturbations of the order parameter will become so
sluggish that the time spent by the system in the vicinity of $T_c$ will
be comparable with the relaxation time scale itself.  The correlation
length corresponding to such a ``freezeout instant'' will set the size of
the regions over which the same vacuum can be selected.  Hence, it will
set the resulting density of the topological defects.  
This prediction (Zurek, 1984; 1985; 1993) has been now borne out
by the experiment (Hendry \etal, 1994; Hendry \etal, 1995; Zurek, 1994).

The aim of this paper is to discuss the dynamics of the non-conserved
order parameter during the course of the phase transitions in the wake of the
``cosmological''  quench experiments which were recently carried out in
superfluid helium (Hendry \etal, 1994; Hendry \etal, 1995) and in
liquid crystals (Chuang \etal, 1991; Yurke, 1995; Bowick \etal, 1994)
and to propose further experiments which can shed a new light on the
heretofore unaddressed aspects of the symmetry breaking.  We shall
focus on second-order phase transitions in superfluids
and in superconductors, which constitute two experimentally
attractive alternative implementations of symmetry breaking phase
transitions in systems with a global and local gauge, respectively.

In the next section we shall briefly review ``static'' (equilibrium)
aspects of symmetry breaking in field theories, in the superfluid
helium, and in the superconductors.  Section~3 will discuss topological
defect formation in superfluid He$^4$, compare theoretical predictions
with the recent experiment, and consider prospects for analogous
experiments in superconductors.  Section~4 will focus on phase
transitions in annular geometry, in both superfluids and
superconductors.  In such geometry, interplay of the symmetry
breaking with the torus topology of the system results in particularly
dramatic predictions.  Section~5 contains a an application of these 
considerations to the cosmological phase transitions. Conclusions are 
summed up in Section 6.

\bigskip
\noindent
\item{\bf 2.}{\bf SPONTANEOUS SYMMETRY BREAKING IN HIGH ENERGY
AND IN LOW TEMPERATURE PHYSICS}

\nobreak
The ability to draw parallels between the high energy and condensed
matter phenomena rests on the similarity of the behavior of free
energy, especially in the vicinity of the phase transition.  In particular,
for the second-order phase transitions, potential contribution to the
free energy is of the Landau-Ginzburg form:
$$
V(\varphi) = \alpha (T - T_c) |\varphi|^2 + {1 \over 2} \beta |\varphi|^4\;,
\eqno (2.1)
$$
where $\varphi$ is the order parameter or a field (see Fig.~1).  The coefficients of
the two forms have a well-prescribed dependence on the relative
temperature:
$$
\epsilon = (T - T_c)/T_c\;. \eqno (2.2)
$$
Thus:
$$\eqalignno{
\alpha &\simeq \alpha' \epsilon\;, &(2.3)\cr
\noalign{\hbox{and}}
\beta &= {\rm const.} &(2.4)\cr
}
$$

In the field theories the form of the potential energy, Eq.~(2.1), is
often simply postulated (although it can be also justified by the
Gaussian approximation at a finite temperature; Aitchison and Hey,
1982).  In the low-temperature (or, more generally, condensed matter)
context, it is, on the other hand, usually derived in the mean field
approximation from the underlying microscopic theory of the system in
question. For instance, in the case for superconductors, the so-called
Gorkov equations provide a link between the microscopic BCS theory
and the Landau-Ginzburg theory (Gorkov, 1959).  In any case, the
parallels between the equilibrium aspects of the phase transitions in
the field theories relevant to high-energy physics and in the effective
field theories emerging in the mean field description of condensed
matter system have been appreciated for quite some time (Aitchison
and Hey, 1982, and references therein).

Symmetry breaking arises when $T < T_c$, that is when the coefficient
$\alpha$ in Eq.~(2.1) becomes negative.  Then the global minimum of
the potential energy given by Eq.~(2.1) moves from $\varphi = 0$ (where
it resides at $T > T_c$) to the finite absolute value given by
$$
\sigma = \sqrt{|\alpha|/\beta}\;. \eqno (2.5)
$$
When $\varphi$ is two dimensional instead of just two alternatives
($+\sigma$ and $-\sigma$), degenerate minima form a circle of radius
$\sigma$.  For 3-D
$\varphi$, a sphere of radius $\sigma$ constitutes the degenerate
manifold of the alternative true vacua.  The depth of the minimum of
$\sigma$ is given by
$$
\Delta V = V(0) - V(\sigma) = \alpha^2/2\beta\;.  \eqno (2.6)
$$

The tension between the long range order which is supposed to set in below 
the critical temperature and the relatively short range over which the choice
of the broken symmetry vacuum can be communicated and in a finite time 
determined by the rate at which the phase transition is taking place is 
responsible for creation of the topological defects. The tension between 
the long range and locality constitutes therefore the primary focus of 
the dynamics of nonequilibrium phase 
transitions which are the subject of this paper. 

\noindent
{\bf Field Theories}

\nobreak
To consider consequences of the symmetry breaking in field theories,
we supplement potential energy, Eq.~(2.1), with a kinetic term and
consider a Lagrangian $L$:
$$
L(\varphi) = \partial_\mu \varphi^\ast \partial^\mu\varphi -
\alpha\varphi^\ast\varphi - {\beta \over 2}
\left(\varphi^\ast\varphi\right)^2\;, \eqno (2.7)
$$
where $\varphi$ is a complex field.  When $\alpha < 0$, potential energy
associated with Eq.~(2.7) has a degenerate minimum---a circle at the
radius $\sigma = \sqrt{|\alpha|/\beta}$.  Thus, the simplest (static,
space-independent) solution of Eq.~(2.7) is given by
$$
\varphi(x) = \sigma e^{i\theta}\;, \eqno (2.8)
$$
where $\theta$ is the fixed phase.  Small perturbations around such a
uniform solution can be considered.  That is, one may investigate the
behavior of
$$
\varphi (x,t) = \sigma + \left(u(x,t) + i v (x,t)\right)/\sqrt{2}\;, \eqno
(2.9)
$$
where $u,v \ll \sigma$.  Above, $u$ is the perturbation of the absolute
value of
$\varphi$, and $v$ is the perturbation of its phase.

When $L$ is expressed in terms of $u$ and $v$ defined by Eq.~(2.9) and
the constant terms are ignored,
$$\eqalign{
L &= {1 \over 2}\left(\partial_\mu u\right)^2 + {1 \over 2}
\left(\partial_\mu v\right)^2 - \beta \sigma^2 u^2 -\cr
&\quad \left({\beta \over \sqrt{2}}\right) \sigma u \left(u^2+ v^2\right)
- \left({\beta \over 8}\right) \left(u^2 + v^2\right)^2\cr} \eqno (2.10)
$$
obtains.  One can regard it as a Lagrangian for two coupled fields, $u$ and
$v$.  Field $u$ has a positive mass given by $\beta \sigma^2 = -\alpha =
|\alpha|$, and $v$---the so-called Goldstone boson---is massless.

Let us now consider a more interesting static solution of Eq.~(2.10).  To
do this, we begin by deriving a dimensionless version of the Lagrangian
$L$, which obtains when $\varphi$, the distances, and the time are
expressed in the ``natural'' units:
$$\eqalignno{
\varphi \to \eta &= \varphi/\sigma &(2.11)\cr
\vec r \to \vec\varrho &= \vec r/\xi &(2.12)\cr
t \to \tilde t &= t/\tau &(2.13)\cr
}$$
where $\sigma$ is given by Eq.~(2.5), while $\xi$ and $\tau$ are the
correlation length and relaxation time and are given by
$$
\xi = \tau = {1 \over \sqrt{\alpha}} \eqno (2.14)
$$
in the convenient cosmological unit system ($\hbar = c = k_B = 1$) we
are employing in this section.  Now a time-independent solution will
have to satisfy
$$
\nabla^2\eta = \left(|\eta|^2 - 1\right)\eta\;. \eqno (2.15)
$$
Apart from the trivial solution $|\eta|^2 = 1$, Eq.~(2.15) has an
axisymmetric solution of the form:
$$
\eta = f(\varrho) \exp (in \phi)\;, \eqno (2.16)
$$
where ($\varrho,\phi,z$) are the cylindrical coordinates.  Above, $n$
must be a whole number---otherwise, $\eta$ could not be single
valued.  The radial part of the above solution is regular near the origin
($f(\varrho) \simeq \varrho^n$ for $\varrho \ll 1$) and approaches the
equilibrium value at large distances ($f^2(\varrho) \simeq 1 -
n^2/\varrho^2$ for $\varrho \gg 1$).  The phase of this solution is
$\theta = n\phi$ on any $\varrho = {\rm const.} > 0$ circle and is
undefined for $\varrho = 0$, where the amplitude of the broken
symmetry phase decreases to zero.

Equation~(2.16) represents, of course, a {\it string}---topological
defect---with the original symmetric vacuum locked out along the axis
of symmetry by the broken symmetry phase in which a different vacuum
is selected along the different radial directions in such a fashion that
the transition to the minimum energy state, $\eta = 1$, is prohibited by
an infinite energy barrier.  The string tension---energy per unit length
of the object defined by Eq.~(2.16)---is logarithmically dependent on
the cutoff in the distance.  This solution was actually considered first
by Ginzburg and Pitaevskii in the context of Landau-Ginzburg analysis
of superfluid helium (Ginzburg and Pitaevskii, 1958), and we shall soon
return to its interpretation as a superfluid vortex.

An even more common and interesting field-theoretic model
corresponds to a Lagrangian:
$$
L = - {1 \over 4}B_{\mu\nu}B^{\mu\nu} +
\left[\left(\partial_\mu + ieA_\mu\right)\varphi^\ast\right]\;
\left[\left(\partial_\mu - ie A_\mu\right)\varphi\right] -
\alpha\varphi^\ast\varphi - {\beta \over 2} (\varphi^\ast\varphi)^2
\eqno (2.17)
$$
with $A_\mu$ a massless gauge boson and
$$
B_{\mu\nu} = \partial_\mu A_\nu - \partial_\nu A_\mu\;. \eqno (2.18)
$$
While the Lagrangian given by Eq.~(2.10) is invariant under a global
gauge symmetry (i.e., redefining the phase $\theta$ globally),
Eq.~(2.17) respects a more physically interesting local gauge
transformation in which the phase is a function of space, that is
$$\eqalignno{
\varphi(x) &\to e^{-i\theta(x)} \varphi(x) &(2.19)\cr
\noalign{\hbox{and}}
A_\mu(x) &\to A_\mu (x) -e^{-1}\partial_\mu\theta (x)\;.&(2.20)\cr
}$$

When the expansion around the local minimum of $|\varphi|=\sigma$ we
have implemented before for the case of the global gauge with the help
of the substitution, Eq.~(2.9), is carried out for the locally gauge
invariant theory of Eq.~(2.17), we will find
$$\eqalign{
L &= -{1 \over 4}B_{\mu\nu}B^{\mu\nu} + e^2 \sigma^2 A_\mu A^\mu +
{1 \over 2} (\partial_\mu u)^2 + {1 \over 2} (\partial_\mu v)^2\cr
&\quad \beta \sigma^2u^2 - \sqrt{2}\, e \sigma A_\mu \partial_\mu v +
\ldots\cr}
\eqno (2.21)
$$
The surprising (and famous) result of symmetry breaking is the term
involving $A_\mu A^\mu$---it appears that the gauge vector field has
acquired a mass (Aitchison and Hey, 1982).

The excitations of the amplitude of $\varphi$ around $\sigma$ are, of
course, still massive.  To see this, we can further simplify Eq.~(2.21) by
fixing the gauge so that $\theta (x)$ is the phase of the original
complex field
$\varphi (x)$.  Then Eq.~(2.9) reduces to
$$
\varphi (x) = \sigma + u (x)/\sqrt{2}\;, \eqno (2.22)
$$
and the Lagrangian given by Eq.~(2.21) becomes
$$\eqalign{
L &= {1 \over 4} B'_{\mu\nu}B'^{\mu\nu} + e^2\sigma^2 A'_\mu A'^\mu +
{1 \over 2} (\partial_\mu u)^2 - \beta \sigma^2 u^2 - {1 \over 8} \beta
u^4\cr &+ {1 \over 2} e^2 (A'_\mu) \left(\sqrt{2}\, \sigma u +
u^2\right)\;.\cr}
\eqno (2.23)
$$
In this form it is apparent that $L$ describes interaction of the massive
vector boson $A'_\mu$ with a real scalar field $u$, the Higgs boson,
with the mass given by
$$
m_H = \beta \sigma^2 = |\alpha|\;. \eqno (2.24)
$$

Vortex solution still exists (Nielsen and Olesen, 1973) for the
Lagrangian given by Eq.~(2.17), but its properties are now different
than those given by the global string of Eq.~(2.16).  We shall consider it
explicitly in a specific gauge below, in the course of the discussion of
flux tubes in type~II superconductors.  Here, let us only anticipate one
big difference:  String tension associated with the local string is finite
(rather than logarithmically divergent as was the case for global
strings).

\noindent
{\bf Superfluid Helium 4}

\nobreak
The classic condensed matter example of a state with nontrivial broken
symmetry phase is superfluid He$^4$.  Phase diagram of helium is
shown in Fig.~2.  Superfluid HeII occupies the low temperature ($T \lap\
2\degr$~K) low pressure ($p \lap\ 25$~atm) corner of the phase
diagram.  Astonishing properties of superfluid helium (Tilley and
Tilley, 1986) are associated with the Bose condensate of He$^4$
atoms.  Below the pressure dependent critical temperature $T_c$
(which is also often called $T_\lambda$ or ``the $\lambda$-line''
because of the behavior of specific heat in its vicinity), a temperature
dependent fraction of atoms condenses out into a quantum fluid
described by a Bose condensate wave function $\Psi$:
$$
\Psi = |\Psi| e^{i\theta}\;. \eqno (2.25)
$$
It is the phase $\theta$ of $\Psi$ which is responsible for many of the
intriguing properties of superfluid helium.  Some of them can be
understood when we assume that $\Psi$ obeys a Schr\"odinger equation
of the form
$$
i\hbar {\partial\Psi \over \partial t} = -{\hbar^2 \over 2 m}
\nabla^2\Psi + \mu \Psi\;,\eqno (2.26)
$$
where $m$ can be taken to be the mass of He$^4$ atom, and $\mu$ is the
chemical potential---the energy gained by the system as a result of
the addition of a single particle at constant volume and entropy.

The equation for superfluid flow can be obtained from the
Schr\"odinger equation (2.26) by computing the rate of change of
$|\Psi|^2$ where this probability density is now interpreted as the
density of the superfluid (Tilley and Tilley, 1986).  Thus, for example,
one can show that the velocity $v$ of the superfluid and the phase
$\theta$ of $\Psi$ are connected with a simple equation:
$$
\vec v = {\hbar \over m} \vec\nabla \theta\;. \eqno (2.27)
$$
The formal connection between superfluid helium and our field-theoretic
considerations of the previous section can be established when it is
assumed that the free energy density can be expanded in powers of
$|\Psi|^2$ (which plays the role of the order parameter) and that it has a
Landau-Ginzburg form:
$$
f(\vec r) = \alpha \left|\Psi (\vec r)\right|^2 + {\beta \over 2}
\left|\Psi (\vec r)\right|^4 + {\hbar^2 \over 2 m} \left|\vec\nabla \Psi
(\vec r)\right|^2\;. \eqno (2.28)
$$
When the chemical potential $\mu$ is evaluated with the help of the
above ansatz, the Schr\"odinger equation (2.26) we have written
before acquires a form:
$$
i\hbar {\partial\Psi \over \partial t} = - {\hbar^2 \over 2m} \nabla^2
\Psi + \alpha \Psi + \beta |\Psi |^2\Psi\;. \eqno (2.29)
$$
This relation is known as the Gross-Pitaevskii equation.  It is generally
treated with suspicion by the condensed matter theorists, who
cite evidence that the so-called time-dependent Ginzburg-Landau (TDGL)
equation (which has, on the left-hand side, a real rather than imaginary
number multiplying the time derivative $\partial\Psi/\partial t$) is a
better model for many systems (including superfluid He$^4$) in the vicinity
of the phase transition (Hohenberg and Halperin, 1977).

Indeed, in the related case of superconductors TDGL equations have been 
rigorously derived from the microscopic theory (Gorkov, 1959). In that case 
the system is definitely overdamped, with the left-hand side of the equation 
for the order parameter dominated by $\Gamma ~ \partial \Psi / \partial t$
where $\Gamma$ is real. Although the microscopic theory of He$^4$ superfluidity
does not exist, the reason for overdamping (which in the superconductors is due 
to the coupling between the super and normal electrons) is certainly also 
present in superfluid helium. Thus, one can expect that approach to equilibrium 
will not be governed by Eq.~(2.29), and this is indeed the case.  The TDGL
equation represents such relaxation quite accurately.  Nevertheless,
Eq.~(2.29) does have some interesting features.  For example, the
dispersion relation which obtains from the Gross-Pitaevskii equation
does have both massless ($\omega \sim k$) and massive ($\omega \sim
k^2$) components, although its form is not reminiscent (Bernstein and
Dodelson, 1991) of the (second sound) phonon-roton dispersion relation
which emerges only when the interatomic forces are taken into account
(Vinen, 1969; Feynman, 1954).

We shall not depend on the Gross-Pitaevskii equation (2.29) in
analyzing details of the rapid phase transition (the subject of the next
section).  It is nevertheless good to write it down and use it to extract
(with the help of the Landau-Ginzburg free energy) several
important pieces of information which turn out to be largely
independent of this (rather suspect) method of their derivation.  The
form of Eq.~(2.29) differs from the equation of motion one would obtain
from the globally gauge invariant Lagrangian, Eq.~(2.7), only in that it
has a first (rather than second) time derivative on the left-hand side.
As a result, when Eq.~(2.29) is re-scaled to assure a dimensionless
form,
$$
i\dot\eta = -\nabla^2\eta + \left(|\eta|^2 - 1\right) \eta\;,  \eqno
(2.30)
$$
the following three important quantities---relaxation time $\tau$,
correlation length $\xi$, and equilibrium density of the superfluid
$\sigma^2$---are employed:
$$\eqalignno{
\tau &= \hbar/|\alpha| = \tau_0/|\epsilon|\;, &(2.31)\cr
\xi&= \hbar/\sqrt{2m|\alpha|} = \xi_0/\sqrt{|\epsilon|}\;, &(2.32)\cr
\noalign{\hbox{and}}
\sigma^2 &= -\alpha/\beta\;. &(2.33)\cr
}$$
Scalings with temperature follow when we adopt the usual ansatz 
[Eqs.~(2.2)-(2.4)] for
the parameters $\alpha$ and $\beta$ (see Ginzburg and Pitaevskii, 1958,
for numerical estimates of $\alpha'$ and $\beta$):
$$\eqalignno{
\alpha &= \alpha' \epsilon\;,\quad \alpha' \cong 10^{-16} [{\rm erg}]\;;
&(2.34)\cr
\beta &= {\rm const.}\;,\quad \beta \cong 4 \cdot 10^{-40} [\hbox{erg
cm}^3]\;, &(2.35)\cr
}$$
where $\epsilon = \left(T - T_c\right)/T_c$, Eq.~(2.2), is the relative
temperature.

When we restrict ourselves to the time-independent solutions of
Eq.~(2.30), we will be led back to Eq.~(2.15) that we have already
considered in the preceding subsection.  In particular, the global string
solution given by Eq.~(2.16) for $n = 1$ is known as the
Ginzburg-Pitaevskii vortex in the superfluid helium context (Ginzburg
and Pitaevskii, 1958) (see Fig.~3).  The flow around the axis of
symmetry is caused by the phase gradient, so that at a distance $r$
away from the core of the vortex, its velocity is given by
$$
v = {\hbar \over m} \cdot {1 \over r}\;. \eqno (2.36)
$$
Vortices with the winding number $n > 1$ are unstable, as
they are  energetically more expensive than the $n = 1$ case.  This can
be seen by computing string tension $\varsigma$, which can be readily
obtained from the kinetic energy of the flow for an arbitrary winding
number $n$ out to some cutoff $L$:
$$\eqalign{
\varsigma &= \int^L_0 2\pi r\, dr \cdot \left(m |\Psi|^2 v^2\right)/2\cr
&\approx n^2 \cdot {\pi\hbar^2 \sigma^2 \over m} \cdot \ln \left({L
\over \xi}\right)\;. \cr} \eqno (2.37)
$$
Clearly, topological constraint (i.e., a certain winding number $N$ within
a closed path at some large radius) can be satisfied at a lesser
energetic expense by $N$ vortices with $n = 1$ each rather than by a
single vortex with $n = N$, for any $N > 1$.

The string tension, Eq.~(2.37), can be re-written in a more suggestive
way:
$$
\varsigma = \left[\pi \xi^2 \cdot \left(\alpha^2/2\beta\right)\right]
\cdot 4 \ln \left({L \over \xi}\right)\;. \eqno (2.38)
$$
This equation [obtained from Eq.~(2.37) in the case of $n = 1$] lends
itself to an intuitively appealing interpretation:  String tension is
approximately proportional to the energy of the symmetric vacuum
which occupies the core (of radius $\xi$) of the vortex (and which is
given by the term in square brackets).

We also note the logarithmic dependence of $\varsigma$ on the
large-scale cutoff $L$.  [The small-scale cutoff at $\xi$ appears
naturally because of the absence of the superfluid condensate at small
radii, where $|\Psi|^2 \sim \left({r \over \xi}\right)^2$, see discussion
below Eq.~(2.16).]

Existence of vortex lines was postulated nearly fifty years ago by
Onsager (1949) and (independently, but somewhat later) by Feynmann
(1954) to allow for superfluid rotation in spite of the requirement of
single-valuedness of the Bose condensate wave-function.  Vortex lines
have since been studied both experimentally and theoretically (Tilley
and Tilley, 1986; Donelly, 1991), although their possible origin in the
``cosmological'' scenario seems to have evaded attention of many of the
low temperature scientists (see, for example, discussion of the vortex
creation during the superfluid phase transition in Donelly, 1991, section 5.7).

In addition to the vortex lines, we shall consider ``cosmological''
implications of persistent superfluid flows in an annular container.  This
phenomenon is again a consequence of the crucial role of the quantum phase of
the Bose condensate in HeII:  When the phase $\theta$ increases by
$2\pi n$ as one follows a closed path along the (large) circumference $C
= 2\pi r$ of the torus containing superfluid He$^4$, the velocity
associated with the gradient of phase will be given by Eq.~(2.27),
$$
v = ({\hbar \over m}) \cdot {2\pi n \over C} = {\hbar \over m} \cdot {n
\over r}\;. \eqno (2.39)
$$
Such flows are also stable (although for a somewhat different reason
than the flow around the vortex line).  Moreover, now $n$ is essentially
arbitrary [that is, until $v$ reaches critical velocities at which
superfluid flow begins to be dissipative (Tilley and Tilley, 1986)].

We end this subsection with a caveat:  Landau-Ginzburg theory is only a
qualitatively correct approximation for superfluid He$^4$.  Therefore,
scaling relations for various quantities with the relative temperature
$\epsilon$ are somewhat different than one would infer from
Eqs.~(2.32)--(2.35).  For instance, measurements show (Ahlers, 1976)
that the correlation length $\xi$ scales as $|\epsilon|^{-\nu}$, where
$\nu = 2/3$ and is consistent with the predictions of the
renormalization group approach (see, e.g., Goldenfeld, 1992).  We shall
pay attention to the consequences of such differences in what follows.
Here, let us only note that the discrepancy between the
Landau-Ginzburg approach and the behavior of superfluid He$^4$ is not
unexpected.  This is because, on the one hand, mean field theory which
would allow one to apply the potential energy model of Eq.~(2.21) to
systems with atomic substructure is valid only when the correlation
length $\xi$ is much larger than the interatomic spacing $a$,
$$
\xi \gg a\;. \eqno (2.40)
$$
On the other hand, considerations based on the Landau-Ginzburg model
are valid only when the effects of thermal fluctuations can be
neglected.  This means that the inequality:
$$
E_G = \xi^3 \cdot \Delta V  > {1 \over 2} k_B T_c\;, \eqno (2.41)
$$
where $\Delta V = \alpha^2/2\beta$, Eq.~(2.6), known as the {\it Ginzburg
condition\/} should be satisfied.

In superfluid He$^4$ the correlation length far away from $T_\lambda$
is of the order of 4--5 \AA ngstroms.  Therefore Eq.~(2.40) would
demand $\epsilon \ll 1$.  But (with the Landau-Ginzburg scaling $\xi
\sim 1/\sqrt{|\epsilon|}$) the left-hand side of Eq.~(2.41) scales as
$\sqrt{|\epsilon|}$, and the inequality is satisfied only approximately
0.5$\degr$K below $T_\lambda$.

Hence, the two conditions, Eq.~(2.40) and Eq.~(2.41), cannot be
simultaneously satisfied.  As a result, the Landau-Ginzburg theory can
be only applied as a qualitative guide to the behavior of superfluid
helium (Tilley and Tilley, 1986).  We shall return to the condition of
Eq.~(2.41) below, as it defines Ginzburg temperature:
$$
\left|\epsilon_G\right| = \left({\beta k_B T_c \over
\xi^3_0 \alpha'^2}\right)^2\;. \eqno (2.42)
$$
This relative temperature characterizes the demarcation line above
which the potential difference between the minimum and the central
peak is sufficiently small to make it easily traversable by the thermal
fluctuations in the correlation length sized regions.

\noindent
{\bf Superconductors}

\nobreak
Superconducting phase transition can be modeled by a
Landau-Ginzburg mean field theoretic model with the free energy
given by (Tilley and Tilley, 1986; Tinkham, 1985; Werthamer, 1969)
$$
F = \alpha |\Psi|^2 + {1 \over 2}  \beta |\Psi|^4 + {1 \over 4m}
\left|\left(-i\hbar \vec\nabla - {2e \over c} \vec A\right) \Psi\right|^2
+ {B^2\over 8\pi} + E_0\;. \eqno (2.43)
$$
Above, $\Psi$ is the order parameter---wave function of the Bose
condensate of Cooper pairs with mass $2m$ and charge $2e$---while
$\vec A$ is the vector potential and
$$
\vec B = \vec\nabla \times \vec A\;. \eqno (2.44)
$$
Energetic contributions due to the applied (external) field are
incorporated in the constant $E_0$.

Equation~(2.43) can be justified rigorously by appealing to
microphysics.  The connection relies on the effective Gorkov equations
(Gorkov, 1959) which can be derived from the
BCS theory (Bardeen, Cooper, and Schrieffer, 1957) of
superconductivity in the $\epsilon \ll 1$ region.  Thus, in contrast to
superfluid He$^4$, the Landau-Ginzburg theory is not just a
phenomenological, qualitative approximation, but a microscopically
justifiable quantitative tool (Tilley and Tilley, 1986; Tinkham, 1985).
This fact will validate and simplify our discussion.  On the other hand,
instead of just one He$^4$, the list of superconducting materials is
rather long, even if we were to ignore the new high-$T_c$ additions.
Moreover, existence of the gauge field adds another characteristic
length scale to the problem.

Thus, by the same token as before, we can define the relaxation time
scale,
$$
\tau = \hbar/|\alpha| = \tau_0/|\epsilon|\;, \eqno (2.45)
$$
where $\tau_0$ can be derived from the Gorkov equation
$$
\tau_0 = {\pi\hbar \over 16k_B T_c} \approx 1.5 \cdot 10^{-12}T_c^{-1}
[s]\;. \eqno (2.47)
$$
Correlation length:
$$
\xi = \hbar/\sqrt{4m|\alpha|} = \xi_0/\sqrt{|\epsilon|} \eqno (2.48)
$$
characterizes variations of the order parameter while London's
penetration depth $\lambda$ is proportional to the correlation length,
$$
\lambda (T) = \kappa\cdot \xi (T)\;, \eqno (2.49)
$$
where $\kappa$ is the temperature-independent constant which
distinguishes between the two kinds of superconductors:  Type~I
(where $\kappa < 1/\sqrt{2}$) and type~II (where $\kappa > 1/\sqrt{2}$).

For us this distinction is critically important, as the vortices can exist
only in type~II superconductors.  There, the axially symmetric
nontrivial solution which is a local minimum of Landau-Ginzburg free
energy, Eq.~(2.43), has the order parameter with the absolute value:
$$
\left(\Psi/\sigma\right) \sim \tanh {br \over \xi}\;, \eqno (2.50)
$$
where $b$ is a constant close to unity.  In the limit of extreme type~II
superconductors ($\kappa \gg 1$), the induction varies as
$$
B(r) \cong {\Phi_0 \over 2\pi\lambda^2} K_0 \left({r \over
\lambda}\right)\;, \eqno (2.51)
$$
where $K_0$ is a zero-order Hankel function of imaginary
argument, and $\Phi_0$ is a flux quantum:
$$
\Phi_0 = hc/2e\;. \eqno (2.52)
$$
Qualitatively, $B(r)$ falls off as $e^{-r/\lambda}$ at large radii ($r >
\lambda$) and can have significant values only in the shell $\xi < r <
\lambda$ where it diverges as $\ln (\lambda/r)$.  This shell disappears
in type~I superconductors (for which $\xi > \sqrt{2}\lambda$).  Hence,
type~I superconductors do not allow for existence of vortices.

String tension of vortex lines is finite, and it is given by
$$
\varsigma \approx ({\Phi_0 \over 4\pi\lambda})^2 \ln \kappa =
{H_c^2 \over 8\pi} \cdot 4\pi \xi^2 \ln \kappa\;, \eqno (2.53)
$$
where $H_c$ is the so-called critical field (which, when applied
externally, expels superconducting phase from the sample).  This
formula is interesting because it demonstrates that the energetic cost
of getting rid of the Bose condensate is given by the energy density
associated with the symmetry breaking, $\Delta V$ of Eq.~(2.6), so that
$$
\Delta V = \alpha^2/2\beta = H^2_c/8\pi\;. \eqno (2.54)
$$
Thus, Eq.~(2.53) implies that the string tension corresponds to the
energy of the condensate displaced by the tube of radius $r$ of normal
metal times of proportionality factor of $4 \cdot \ln\kappa$.  Similarly,
one can check that $\varsigma$ is of the order of the energy of the
field configuration associated with the flux line.

As was the case for superfluids, one can also consider loops of
superconductors in an empty space.  Single-valuedness of the order
parameter requires that the phase accumulated along the
circumference of the loop must correspond to an integer multiple of
$2\pi$.  Now, however---in contrast to the case of He$^4$---there are
two sources of the phase gradient:  In addition to the velocity $\vec v$
of the ``superfluid'' of Bose-condensed Cooper pairs, there is also the
vector potential.  Consequently, the quantization condition is
$$
{c \over 2e} \oint_C \left(2m\vec v + {2e\vec A \over c}\right)
d\vec l = n\Phi\;. \eqno (2.55)
$$
In the limit in which the superconducting loop is wider than the Landau
penetration depth $\lambda$ (so that there are closed paths $C$ along
which the first term containing $\vec v$ is negligible), Eq.~(2.55) implies
quantization of the flux in units of $\Phi_0$.  This need not always be the
case, as we shall find out while analyzing phase transitions in the annular
geometry.  Then $\lambda (T)$ becomes large, and sufficiently near
$T_c$ the resulting flux will be quantized in units smaller than $\Phi_0$,
with the velocity term making up the difference of the phase of the
Bose condensate.

The reliability of the Landau-Ginzburg theory for superconductors is
due to the significant values of $\xi_0$ (typical $\xi_0 \gap\
1000\,\hbox{\AA}$, more than two orders of magnitude larger than in
He$^4$).  Thus, the Ginzburg energy which is approximately given by
$$
E_G = \sqrt{\epsilon} \cdot \xi^3_0 H_c^2(T = 0)/8\pi \eqno (2.56)
$$
[see Eq.~(2.54)] is typically much larger than the corresponding
quantity for the same $\epsilon$ in superfluid helium.  To get an order
of magnitude estimate of the relative Ginzburg temperature
$\epsilon_G$, we compute
$$
\epsilon_G = 1.2 \times 10^{-5}
\left({8\pi k_B T \over \left(\xi_0/1000\,\hbox{\AA}\right)^3
\left(H_c/100\, Oe\right)^2}\right)^2\;. \eqno (2.57)
$$
Thus, thermal fluctuations large enough to alter the configuration of the
field are found only in the immediate vicinity of the critical
temperature, even when a rather modest value for $H_c$ (which can be
$\sim 10^3Oe$) and $\xi_0$ (which is typically closer to a few
thousand \AA ngstroms in type~I superconductors) is adopted.

It is perhaps useful to point out that the above estimates do not apply
to the high temperature superconductors, which have small $\xi_0$ and
much more modest critical fields $H_c$ (Salamon, 1988).

\noindent
{\bf Other Systems}

\nobreak
The above two condensed matter phase transitions---the superfluid
He$^4$ and super\-conductors---do not exhaust the list of the
experimental possibilities, and the aim of this section is to point
out---very briefly---that there are other systems in which
topological defects in a nonconserved order parameter can be (or,
indeed, have been) studied with the ``cosmological'' motivations in
mind.

{\bf Liquid crystals} were already mentioned in this context. They were
the first system in which a version of the experiment suggested for
superfluid He$^4$ was carried out 
(Chuang \etal, 1991; Bowick \etal, 1994). 
Nematic liquid crystals consist of rod-like molecules (see for example
de Gennes, 1974). Above the phase transition temperature these molecules
are oriented randomly, but as the temperature falls they tend to align.
It is not difficult to imagine how this local order may lead to topological
defects.

Formaly, the order parameter in the liquid crystal is given by the director 
field $\vec n(\vec r)$, which is just the local orientation of the rods. 
By definition $| \vec n (\vec r)| = 1$. Moreover, since the molecules can 
be flipped by the angle $\pi$ without changing the physical configuration
of the system, $\vec n (\vec r) = - \vec n (\vec r)$. The vacuum manifold 
is therefore a two-sphere with the antipodal points identified. 

Topologically stable defects which are expected in such a symmetry breaking 
scheme include strings (also known as disclination lines) and a monopole. 
Rod - like molecules are arranged in a ``hedgehog'' configuration around 
a monopole. And there are several distinct kinds of linear defects. Rods can, 
for example, simply point away from the core of the string. In that case, 
when the core is circumnavigated the director field changes its orientation
by $2\pi$. But this is not the only possibility. Topologicaly stable defect
can exist also when the director field changes its orientation by only
$\pi$, or by as much as $3\pi$. This is because $\vec n (\vec r)$ and
$- \vec n (\vec r)$ are identical. These topological defects can transform
into each other in course of interactions.

Phase transition into the nematic liquid crystal is of the {\it first} order.
Following such a phase transition large densities of topological defects 
(especially the string-like disclination lines) were detected and studied
(Chuang \etal, 1991; Bowick \etal, 1994). 
Disclination lines are visible under a microscope. Therefore, their 
scaling dynamics can be followed in detail. The disadvantage (from our point
of view) is that the relevant phase transition is (weakly) first order. Thus,
the most intriguing first stages of the development of the new broken symmetry
phase can be understood in the relatively straightforward terms of the
traditional nucleation paradigm, and do not shed any new light on the behavior
of the order parameter in course of the second order critical dynamics.

{\bf Superfluid He$^3$} has even richer collection of topological defects.
This is because the elementary particles which form the Bose condensate
are pairs of He$^3$ atoms, and, as such, have a nontrivial internal structure.
This results in additional degrees of freedom which are accessible even at
the very low (miliKelvin) temperatures in which He$^3$ undergoes the superfluid
phase transition. These additional degrees of freedom are associated with
the non-zero spin of the Cooper pair (S=1) and with the orbital angular
momentum of the Cooper pair around its center of mass. In the limit where 
the spin and angular momenta of the Cooper pairs are strongly correlated, 
a liquid crystal like magnetic ordering will emerge. Thus, in addition to
the broken gauge symmetry (present already in He$^4$) rotational symmetry
of both spin and orbital degrees of freedom can be broken. 
Consequently, the order parameter of superfluid He$^3$ is described by
macroscopic wave functions like in He$^4$, but with {\it nine} complex
amplitudes. Moreover, there are two different phases of He$^3$ superfluid:
The quasi-isotropic phase B, which corresponds to Cooper pairs with no total
(that is, spin plus orbital) angular momentum, as well as the anisotropic
phase A, for which total angular momentum is non-zero when projected along some
direction.

As a consequence of the rich structure of the order parameter,
varied and interesting topological defects occur
in both of the superfluid phases of He$^3$ (Volovik, 1992). Moreover one He$^3$
version of the quench experiment has been considered already some time ago
(Mineev, Salomaa, and Lounasmaa, 1987; Salomaa, 1987; Salomaa and
Volovik, 1987), and, as we shall discuss shortly, another one has been recently
carried out (Ruutu \etal, 1996). 

Helium~3 has both advantages and disadvantages when compared with He$^4$.
It is experimentaly more difficult to work with, and the variety of the
topological defects may complicate the interpretation of the experimental 
results. Moreover, its correlation length is orders of magnitude larger than 
in He$^4$, so the density of the defects is expected to be significantly 
less. On the other hand, large correlation length implies that the 
Landau-Ginzburg theory is a much better model, so the analogy with the field
theories is better justified. Moreover, vortices can be detected "one by one"
using nucler magnetic resonance along with some ither clever experimental
techniques (Ruutu \etal, 1996). Hence, lower initial vortex line density
may turn out not to be a problem.

Last but not least, phase diagram of He$^3$ allowes for a variety 
of quench trajectories, some of which raise the possibility of addressing 
other interesting issues such as the creation of topological defects in
tuneably weak first order phase transformations, or even in a quick 
succession of the second and first order phase transitions.

\bigskip


\mag=\magstep1
\hsize 6.5truein
\vsize 8.9truein
\lefthyphenmin=3
\clubpenalty 10000
\brokenpenalty 10000
\displaywidowpenalty 10000
\pretolerance 1000

\def\etal{{\it et~al.}}
\def\ul#1{$\underline{\smash{\vphantom{y}\hbox{#1}}}$}
\def\lap{\hbox{~{\lower -2.5pt\hbox{$<$}}\hskip -8pt\raise
-2.5pt\hbox{$\sim$}}}
\def\gap{\hbox{~{\lower -2.5pt\hbox{$>$}}\hskip -8pt\raise
-2.5pt\hbox{$\sim$}}}
\def\degr{^\circ}

\baselineskip 20pt

\noindent
{\bf 3.\quad QUENCHING OUT VORTEX LINES}

\nobreak
The analogy between the static, equilibrium properties of the field
theories and condensed matter systems implies---as we have seen it
in the preceding section---existence of similar kinds of
topological defects.   The purpose of this section is to employ this
analogy to devise laboratory experiments which would allow one to
study defect formation in the course of rapid phase transitions.  I
shall focus on the condensed matter systems with non-conserved order
parameter only and consider defect
formation starting with the simplest case---superfluid helium.  After
a discussion of the freeze-out scenario, I shall go on to consider its
consequences for the specific experiment already carried out by the
Lancaster University group of Peter McClintock and his colleagues
(Hendry \etal, 1994; Hendry \etal, 1995).  The
theoretical analysis will be repeated for the even more interesting
(local gauge) case of a rapid phase transition to a superconductor, although it
will be pointed out that the main obstacle for a successful study of a
superconducting analog of the cosmological scenario is the difficulty
of rapid detection of the vortices (flux lines).  Thus, discussion of a
more experimentally attractive (although more theoretically
challenging) version of the superconducting
experiment---which involves locking out magnetic flux in a
loop---will be delayed to Section~4.  

\noindent
{\bf Quench Into Bulk Superfluid}

\nobreak
Superfluid He$^4$ has a number of experimental advantages which
make it---in spite of the theoretical complications we have hinted at
in the previous section---perhaps an ideal laboratory to study
dynamics of phase transitions in systems with global symmetry.
Thus, for example, one can obtain samples of He$^4$ of almost
unparalleled purity.  Furthermore, He$^4$ can accommodate
only one kind of defect---vortex lines---simplifying the
analysis.  Moreover, detection of vortex lines---while not as
straightforward as it is for defects in liquid crystals, where they are
visible almost with a naked eye (Chuang \etal, 1991;
Bowick \etal, 1994; Yurke, 1995)---is also relatively simple, as it
relies on the well-tested techniques of second sound attenuation
(Donnelly, 1991).  We have already mentioned that the superfluid
transition is of the second order.  And last, but definitely not least,
phase transition in He$^4$ can be induced on a dynamical time scale by
rapid decompression.  Therefore, the critical temperature
$T_\lambda$ will be reached throughout the bulk of the He$^4$ sample
on a very short time scale limited by the first sound travel time
($\sim 220$~m/s in the vicinity of the critical temperature
$T_\lambda$).  This is orders of magnitude larger than the (second
sound) velocity defined by the ratio of the correlation length and
relaxation time scale:
$$
s = \xi/\tau = \left({\xi_0 \over \tau_0}\right)\sqrt{|\epsilon|}\;,
\eqno (3.1)
$$
which limits the speed with which different regions of the emerging
superfluid become correlated.  Above, $\xi_0/\tau_0$ evaluates to
approximately 70~m/s but second sound reaches this velocity only
quite far from
$T_\lambda$.  In the vicinity of $T_\lambda$ $s$ is much less, so that
the ``second sound horizon'' can be quite a bit smaller than the size of
the sample. Second sound gives the rate of propagation of perturbations of
the density perturbations of the superfluid. Hence, it also limits the rate
at which the order parameter will be able to adjust. In this sense, one 
is able to reproduce the ``acausal''
nature of the cosmological phase transitions in the superfluid He$^4$
better than in the other systems, especially where the transport of
heat is involved (as is the case in liquid crystals and will likely be the
case in superconductors).

The schematic quench trajectory (which proceeds along the isentrope) is
shown in  Fig.~4.  Figure~5 illustrates the transformation of the
effective (Landau-Ginzburg) free energy, as well as its hypothetical
effect on the configurations of the order parameter.  The aim of this
section is to estimate the density of the vortex lines left behind by
the quench.  To do this, we shall rely on the {\it critical slowing
down}---the behavior of the relaxation time scale
$\tau$ in the vicinity of $T_\lambda$.  As we have seen in the previous
section (and as is confirmed by the experiments), $\tau \sim
1/|\epsilon|$.  Thus, in the course of the quench the time scale on which
the order parameter can adjust to the new thermodynamic parameters
(especially to the new value of $\epsilon$!) is becoming very long in
the vicinity of the critical temperature.  As a result, two regimes
can be distinguished: (i)~Sufficiently far from $T_\lambda$ the
relaxation time scale
$\tau$ is much smaller than the time on which the quench is
proceeding.  In this {\it adiabatic\/} regime order parameter will be
characterized by an equilibrium configuration with the correlation
length
$\xi$ determined (to an excellent approximation) by the
instantaneous value of $\epsilon$.  By contrast, very near
$T_\lambda$ the equilibrium relaxation time scale will be much larger
than the time spent by the system with the corresponding value of
$\epsilon$.  As a consequence, we can define:  (ii)~The {\it impulse\/}
region, in which $\tau$ is so large (because of the critical slowing
down) that the configuration of the order parameter will be
in effect immobilized on the timescale of interest.

The boundary between these two regimes will occur at the freeze-out
time $\hat t$.  We can compute it by assuming that, in the vicinity of
$T_\lambda$, the relative temperature $\epsilon$ is approximately
proportional to time:
$$
\epsilon = t/\tau_Q\;. \eqno (3.2)
$$
{\it Quench timescale\/} $\tau_Q$ can be controlled by the rate at
which the pressure is lowered.  Critical temperature is reached at $t
= 0$.

Freeze-out time is then set by the equality:
$$
\tau (\hat t) = \hat t\;, \eqno (3.3)
$$
which, given Eq.~(3.2) above, as well as $\tau = \tau_0/|\epsilon|$,
yields:
$$
\hat t = \sqrt{\tau_0\tau_Q}\;. \eqno (3.4)
$$
Consequently, the transition between the adiabatic and impulse
regimes occurs (twice) during the quench at the relative temperature:
$$
\hat\epsilon = \epsilon (\hat t) = \sqrt{\tau_0/\tau_Q}\;. \eqno (3.5)
$$
Here, $\tau_0 \simeq \hbar/\alpha' \simeq 10^{-11}$~s, with
more careful estimates yielding $\tau_0 = 8.5 \cdot 10^{-12}$~s.  The
fluctuating configuration of the order parameter is frozen out at
$\hat\epsilon > 0$, above $T_\lambda$.  When the dynamics are
``restarted'' again with $\hat\epsilon < 0$, the parameter $\alpha$ is
also negative, which means that the symmetry breaking has already
occurred and topological defects are ``frozen out.''

The characteristic correlation length $\hat\xi$ corresponding to
$\hat t$ and $\hat\epsilon$ above will decide the (initial) density of
the vortex lines through the approximate formula:
$$
\ell_0 = \hat\ell \approx \hat\xi^{-2}\;, \eqno (3.6)
$$
which is based on the idea that typically a piece of the vortex line
will fall within a $\hat\xi$-sized domain (Kibble, 1976; 1980; Vilenkin,
1981).

Correlation length diverges near the phase transition temperatures:
$$
\xi (\epsilon) = \xi_0/|\epsilon|^\nu\;. \eqno (3.7)
$$
In the mean field theory $\nu = 1/2$ [see Eq.~(2.32)] and $\xi_0 =
5.6$~\AA\ provide an acceptable fit to the data.  However,
experiments seem to point to a better accord with the
renormalization group prediction for $\nu = 2/3$ and a somewhat
different $\xi_0 = 4$~\AA\ to go with it (Ahlers, 1976;
Goldenfeld, 1992).  Consequently, the estimated
value of the initial vortex line density is given by
$$
\ell_0 \approx \xi^{-2}_0 \left({\tau_0 \over \tau_Q}\right)^\nu\;.
\eqno (3.8)
$$
The first obvious remark is that---because of the smallness of
$\xi_0$---even for relatively slow quench time scales the anticipated
density of vortex lines is enormous.  The second comment is that this
very large initial vorticity is likely to disappear on a rather rapid
time scale.  This is because the density of vortex lines decays
approximately as (Vinen, 1957):
$$
{d\ell \over dt} = -\chi {\hbar \over m} \ell^2 = -\gamma\ell^2\;,
\eqno (3.9)
$$
where the Vinen parameter $\chi$ is a dimensionless constant.  Thus,
$\ell$ will decrease rather quickly, with the time-dependent density
given by
$$
\ell = {\ell_0 \over 1 + \gamma\ell_0 t}\;. \eqno (3.10)
$$
This process is also reminiscent of the dynamics of cosmic string
network, although absence of cosmological expansion, the
differences in the equation of motion of vortex lines, etc., make this
a rather distant analogy.

Our derivation of the preferred scale $\hat\xi$ and the resulting
initial density of topological defects can be briefly summed up by
noting that we are looking for the instant when the quench becomes
effectively instantaneous.  The correlation length at that instant will
be obviously ``frozen out'' and will set the initial scale of the vortex
line network in the regime where the symmetry of the order
parameter is broken.

There are two other ways of deriving the estimate of the
characteristic scale in the course of second-order phase
transformations which appeal to somewhat different physical input
but lead to the same estimate of $\hat\xi$.  We shall consider them
now as they may be of heuristic value to the readers.

To describe the first of them, we compare the velocity with which
the correlation length would have to increase to always maintain its
equilibrium value,
$$
v_\xi = {d\xi (t)\over dt} = (d\xi/d\epsilon)\dot\epsilon\;, \eqno
(3.11)
$$
with the velocity of the second sound or, more generally, with the
speed of propagation of the perturbations of the order parameter,
$$
s = \xi (\hat\epsilon)/\tau(\hat\epsilon)\;. \eqno (3.12)
$$
Simple algebra shows that $v_\xi$ would have to increase faster than
$s$ when $\epsilon$ is smaller than the characteristic value
$\hat\epsilon$ which satisfies equation:
$$
v_\xi (\hat\epsilon) = s(\hat\epsilon)\;, \eqno (3.13)
$$
or
$$
\xi_0 \hat\epsilon^{-3/2}/2\tau_Q \approx \xi_0
\hat\epsilon^{1/2}/\tau_0\;. \eqno (3.14)
$$
This last equality leads to Eq.~(3.5), providing we ignore numerical
corrections of order unity.  The physical implication of this way of
obtaining $\hat\epsilon$ and $\hat\xi$ are straightforward:  $\hat\xi$
is as large as the dynamics of the order parameter (characterized by
$s$) and the quench timescale $\tau_Q$ allow it to become.

Freezeout temperature $\hat\epsilon$ can be also obtained by
comparing $\xi$ with the size of the sonic horizon $h$---the
distance over which perturbations of the order parameter can
propagate in the course of a quench.  This distance is set by
$$
h = \int\limits^t_0 s(t)\,dt \simeq s(\epsilon) \cdot \epsilon \cdot
\tau_Q\;. \eqno (3.15)
$$
The system will emerge in the broken symmetry phase with the
correlation length no larger than the sonic horizon.  This will be
associated with the relative temperature given by
$$
h(\hat\epsilon) = \xi (\hat\epsilon)\;, \eqno (3.16)
$$
which, again, after some straightforward algebra, leads (approximately)
to Eq.~(3.5).

The three above arguments lead to the same conclusion.  This is because they 
are essentially equivalent.  The only difference between them is the manner 
in which they all appeal to the same concept of two different regimes of
the quench---(i)~slow, nearly adiabatic for $|\epsilon|>\hat\epsilon$
and (ii)~almost instantaneous, nearly impulse for $|\epsilon| <
\hat\epsilon$---by respectively comparing timescales, velocities, and distances
characterizing dynamics of the order parameter in the vicinity of the
critical temperature $T_c$.

It may perhaps also be useful to note that the Vinen equation,
Eq.~(3.9), can be reexpressed as an equation for the coherence scale
of the order parameter phase, $D$, by simply defining
$$
D (t) \cong 1/\sqrt{\ell (t)}\;. \eqno (3.17)
$$
$D(t)$ is the size of the domains over which the order parameter has
similar phase.  Using Eqs.~(3.9) and~(3.17), one easily obtains
$$
2 {dD \over dt} = \gamma/D\;. \eqno (3.18)
$$
This is, of course, readily transformed into
$$
{d \over dt} (D^2) = \gamma\;, \eqno (3.19)
$$
which immediately yields the solution
$$
D(t) = D(0) \cdot \sqrt{\gamma t}\;. \eqno (3.20)
$$
This rewriting of the equation for the decay of vortex lines in
superfluid helium in the language of domain size growth offers a
different perspective on the evolution of the order parameter in the
post-quench period.  It also establishes a connection between the
evolution of the vortex line tangle in He$^4$ (see Donnelly, 1991) and
the more general problem of evolution of the domain size in the
systems with nonconserved order parameter.  There, on the basis of
quite general considerations, one can establish Eq.~(3.20) [except for
the case of vortices/monopoles in two dimensions, when $D(t) \sim
\left(t/\ln t\right)^{1/2}$] (Bray, 1994).

\noindent
{\bf Activation Mechanism}

\nobreak
The process we have ignored so far in the discussion---thermal
activation---was initially believed to be essential in generating
topological defects (Kibble, 1976).  As we shall see below, this
expectation does not seem to be borne out in the case of the
superfluid phase transition in He$^4$ [nor was it anticipated in the
original experimental proposal (Zurek, 1984, 1985, 1993)].  To see why
vortex line density is unlikely to reach values mandated by the
thermal activation processes at the Ginzburg temperature, we
consider the initial vortex line network laid down at $\hat t$ by the
freezeout of the order parameter.  Approximately 70\% of the vortex
lines will be in a form of a long string, with only the remaining 30\% in
the form of loops, the smallest of which have size $\hat\xi$. This follows 
from the numerical simulations (Vachaspati and Vilenkin, 1984), 
and approximately coincides with the fraction of nonintersecting 
random walks, although the precise numbers depend on the lattice
as well as on the distribution of the sizes of domains in a way which is not
yet understood (Hindmarsh and Kibble, 1995).  Let us
now suppose that---as will be typically the case---we have stopped 
the quench at some temperature $T$ below the freeze-out temperature
$\hat T = t (\hat t)$, so that the original scale $\hat\xi$ of the defect
network substantialy exceeds $\xi (T)$.  Thermal fluctuations will generate
loops of size $\xi$, which will be typically smaller (much smaller)
than the pre-existing structures.  In particular, the long string can be
modified due to the small-scale warping by the addition of
perturbations on the scale $\xi$; but this process has an essentially
diffusive character, so it will proceed rather slowly.  By contrast, the
long string has a total energy which is much larger than the Ginzburg
energy $E_G$, Eq.~(2.41):
$$
E_{\cal V} = E_G \cdot \left(\ell/\xi\right) \cdot {\cal V} \approx
\left(\xi/\hat \xi\right)^2 \cdot \left(\alpha^2/2\beta\right) \cdot
{\cal V}\;. \eqno (3.21)
$$
Above, ${\cal V}$ is the volume of the container.  This equation is easy
to understand---it can be readily rewritten as
$$
E_{\cal V} = {\cal L} \cdot \xi^2 \cdot \left(\alpha^2/2\beta\right)
\sim {\cal L} \cdot \varsigma\;, \eqno (3.22)
$$
where ${\cal L}$ stands for the total length of the string.  Clearly, for
any macroscopic-sized volume ${\cal V}$ of the superfluid, energy
$E_{\cal V}$ will be much in excess of the thermal energy.
Consequently, while thermal activation energy may suffice to
generate small-scale ringlets of the vortex line, it is unlikely to
elongate the string.  Indeed, if anything, the long string may begin to
straighten before the quench is completed.  This process will
counteract---and will most likely even outweigh---thermal activation events
which would act to increase vortex line density.

Consequently, it is plausible that the dominant vortex line
formation process in the course of a quench will be the freezeout
of the order parameter (Zurek, 1984, 1985, 1993).

\noindent
{\bf Comparison with the He$^4$ Experiment}

\nobreak
In the recent experimental implementation (Hendry \etal, 1994, 
Hendry \etal, 1995) of the cosmologically inspired vortex line formation 
scenario (Zurek, 1984, 1985, 1993), McClintock and his colleagues of 
the Lancaster University have carried out the quench through the $\lambda$-line,
approximately along the isentropic trajectories shown in Fig.~6.  As anticipated
in the ``freezout'' scenario of defect formation, this led to a
copious defect production with the initial line density, estimated to
be
$$
\ell_i \gap\ 10^{13}\, {\rm m}^{-2} = 10^9\,{\rm cm}^{-2} \eqno
(3.23)
$$
in the quench which crossed the $\lambda$-line (Hendry \etal, 1995).
This is certainly a very large vortex line density.  If a comparable
density were to be generated by rotation of the ($\sim 1$~cm
diameter) sample they have used, angular velocities of $\sim 4 \cdot
10^5$ radians per second would be required.

It is possible to estimate the quench time scale $\tau_Q$ from the
change of relative temperature ($\Delta\epsilon \sim 0.1$) and from
the time ($\Delta t \sim 3$~ms) it took to complete the pressure drop:
$$
\tau_Q \cong \Delta t/\Delta\epsilon \cong 30\,{\rm ms}\;. \eqno
(3.24)
$$
Using this estimate of the quench time scale in the equation~(3.8) we
have just derived, we are led to predict
$$
\ell_{LG}[m^{-2}] \simeq 3 \cdot 10^{13}/\left(\tau_Q/100\,{\rm
ms}\right)^{1/2}\;, \eqno (3.25)
$$
when Landau-Ginzburg scaling is employed.  A somewhat more modest
$$
\ell_{RG}[m^{-2}] \simeq 1.2 \cdot 10^{12}/\left(\tau_Q/100\,{\rm
ms}\right)^{2/3} \eqno (3.26)
$$
obtains for the renormalization group scaling (Zurek, 1995).  As it
was already pointed out in the commentary (Zurek, 1994) on Hendry
\etal\ (1994) [where the estimate was evaluated from Eq.~(3.26) for
$\tau_Q = 30$~ms and compared with the then available lower bound
of $\ell_i
\gap\ 10^{11}$~m$^{-2}$], the agreement between the
(order-of-magnitude) estimates offered by the theory and dramatic
experimental results is impressive.  Of course, more could be done
on both theoretical and experimental fronts.  Indeed, progress on the
experimental front may be more rapid (for example, if the proposals
listed in Hendry \etal, 1995, are implemented) than in the theoretical
treatment of the problem (since microphysical theory of
superfluidity in He$^4$ is still missing).

A few remarks on both the experimental side (where I will largely
echo discussion of Hendry \etal, 1995) as well as on theoretical
aspects of the treatment are nevertheless in order.  In addition to:
(1)~the quench through the $\lambda$-line, McClintock and his
colleagues have carried out two other types of quenches;
(2)~quenches far away from
$T_\lambda$; and (3)~quenches starting just below (few milli-Kelvins)
the phase transition temperature.  As we have already noted, copious
production of vortex lines was detected in type~(1) quenches.  By
contrast, no detectable vortex line production was detected
in quenches of type~(2) which never approached $T_\lambda$.  This is
reassuring, as it shows that vortex line creation can occur only where
the transition crosses the $\lambda$-line and freezes-out the
pre-existing fluctuations of the order parameter.  However,
quenches of type~(3) did create detectable vorticity, although in the
amounts which are significantly lower than for the $\lambda$-line
crossing quenches of type~(1).

This is an intriguing observation.  At first, one might be tempted to
appeal to thermally generated seed vortices which could exist in the
``Ginzburg'' regime, but---for the reasons I have already outlined---it
seems unlikely that long (i.e., more than a few correlation length in
diameter) vortex lines may exist a few milli-Kelvin below $T_\lambda$.

If vortices detected following the type~(3) quench from just below
the
$\lambda$-line are not generated by thermal fluctuations alone, what
else can they be?  And why are they detected in type~(3) quenches
but do not appear when the quench starts far from the $\lambda$-line?  
I believe the crucial clue to the interpretation of
this phenomenon may be the remark of Hendry \etal\ (1995),
which attributes appearance of these extraneous vortices to the flows
generated in He$^4$  by the pressure quench.  The difference in the
generated vorticity in type~(2) and in type~(3) quenches may be due
to the different cost---measured in the vortex line tension
$\varsigma$, Eqs.~(2.37) and (2.38)---of creating topological defects.
Thus, if the same energy $\Delta E_{\rm STIR}$ were available in the
stirred-up superfluid,
$$
\ell_{\rm STIR} \sim \Delta E_{\rm STIR}/\varsigma \eqno (3.27)
$$
of vortex line length could be created.  This simple reasoning would
predict that the stirred-up vortex line density should scale (to the
leading order) as
$$
\ell_{\rm STIR} \sim 1/\rho_s\;, \eqno (3.28)
$$
where we have recognized that the vortex line string tension
$\varsigma$ falls off with $\rho_s$ (and where we have ignored
logarithmic terms; if they were taken into account, one would be led
to the equation:  $\ell_{\rm STIR} \cdot \rho_s \sim \left(-\ln
\ell_{\rm STIR} \xi^2\right)^{-1}$, where both the superfluid density
and the correlation length $\xi$ are evaluated near the initial point
on the quench trajectory).

This simple discussion ignores many effects which could be relevant,
but it does have an advantage of being testable:
Quenches along the trajectories which differ only in the initial,
near-$T_\lambda$ points, should allow one to verify (or prove wrong)
Eq.~(3.28) above, perhaps with the logarithmic corrections.

This discussion leads us to one further remark about the vortex line
density generated in quenches which do cross $\lambda$-line.  It is
conceivable that the vorticity generated by the cosmological
mechanism in the course of the quench may be further amplified by
the inadvertent stirring.  If this were indeed the case [and discussion
in Hendry
\etal\ (1995) does admit such possibility], then the direct comparison
of the predictions of our discussion [e.g., Eq.~(3.8)] and of the
experimental results must be taken with a grain of salt.

It would be therefore quite important to have---in addition to the
lower bound given by Hendry \etal, 1995 [Eq.~(3.23)]---to also have
an experimental upper bound.  In view of the rapid evolution of the
vortex line density [Eq.~(3.9)], it might be useful to attempt setting a
thermodynamic upper limit on $\ell_i$.   After all, decay of the vortex
line density will convert string tension energy into the heat
deposited in the superfluid, which should allow one---especially for
$\ell_i \gg 10^{13}$~m$^{-2}$---to set limits on the initial $\ell_i$,
simply by monitoring the temperature of the superfluid.  The
anticipated total temperature increase as a result of decay of
the vortex line density $\ell$ can be estimated:
$$
\Delta T = \ell \varsigma/{\cal Q}\;. \eqno (3.29)
$$
Above, ${\cal Q}$ is the appropriate specific heat of He$^4$ at the
temperature at which the decay of the network is taking place.
Whether detecting corresponding temperature differences resulting
from Eq.~(3.19) may be possible on the scale rapid enough and with
the accuracy sufficient to make the results useful is, of course, a
rather difficult experimental question.\footnote*{Using Eq.~(3.29)
above and a remark in Section~3 of Hendry \etal\ (1995) that ``Fast
expansions seem to follow the quasistatic ones to good
approximation, but, from a common starting temperature, reach a
final temperature which is typically several mK higher'' we can set an
upper limit on $\ell$:
$$
\ell_{15} [m^{-2}] < 0.7 \Delta T [mK]\;.
$$
Thus $\ell_{15} \lap 1$ (or $\ell \lap 10^{15}$~m$^{-2}$) is a plausible
upper limit on the vortex line density.  Moreover in veiw of the
inadvertant ``stirring'' in the course of the quench reported by
Hendry \etal\ (1995), it might be a generous overestimate of the
vortex line density attributable to ``pure'' quench.}

If one were certain that the only way in which the quench could deposit
heat in the superfluid was by dissipation of the string tension energy
locked in the vortex line network generated in the course of the
quench, one could amplify the effect by traversing the
$\lambda$-line many times.  Then the accumulated heat would be due
to the small departures from the isentrope which are caused by
depositing some of the energy in the ordered configurations of the
Bose condensate, which---following each traversal of the
$\lambda$-line---are dissipated in the destruction of the vortex line
network.  Among the assumptions which are made here are an
excellent isolation of the sample as well as a sensible guess that the
energy locked in the vortex network after each quench comes from
the mechanical energy available in this nonequilibrium process
rather than from thermal fluctuations (which, again, relates to the
difference between freeze-out and thermal activation).

In summary, one should stress that all of the caveats listed above do
not weaken the main conclusion of Hendry \etal\ (1994 and 1995).
Very abundant vortex line creation is observed in quenches which
cross the
$\lambda$-line.  Estimates based on the application of the
cosmological scenario of topological defect creation are consistent
with the detected vortex line density, providing that one relies on
the idea of the freeze-out time $\hat t$ rather than on thermal
activation.  Indeed, the final points of all the quenches are well
within the Ginzburg region of He$^4$, which would imply---if one
were to rely on the vortex line creation by thermal
activation---that the final vortex line density should not decay but,
rather, that it should reach values set by the correlation length
associated with the end-of-quench point.  This is clearly not the
case.  Further experimental study of the freeze-out mechanism should
involve limiting the contribution of ``stirring up'' of the superfluid in
the course of the quench as well as varying the quench time scale to
verify Eq.~(3.8).  This course of research has been already mapped out
by the Lancaster University group (Hendry \etal, 1995).

\noindent
{\bf Quenching out Defects in Superfluid He$^3$}

\nobreak
The first demonstration of defects formation in course of the phase transition
into the B phase of superfluid He$^3$ has been just reported by the Helsinki 
group and their collaborators (Ruutu \etal, 1996). The transition ocurrs in 
a cigar-shaped, 100 $\mu$m sized volume in a rotating container of the B-phase
of He$^3$, which is heated above the thermal temperature by the decay products
of a thermal neutron. As this volume cools and re-enters the broken symmetry 
phase, string-like defects form with the density which is consistent with the
freezeout distance $\hat \xi = \xi_0 (\tau_Q/\tau_0)^{1/4}$ of approximately
1 $\mu$m. Ginzburg relative temperature for the superfluid He$^3$ is very small,
and corresponds to the size of the coherence length much larger than the volume
of the ``cigar'' in which the phase transition occurs. Thus, again, the result
of the experiment is inconsistent with the activation mechanism, but appears to 
conform with the freezeout scenario.

The superfluid is slowly rotating. As a result, sufficiently large vortex loops 
generated in the small volume that underwent the phase transition 
are stretched by the Magnus force. This eventualy tends to transform them 
into rectilinear vortex lines parallel to the axis of rotation.
They are subsequently pulled into the center of the container, where they can 
be detected by NMR. The distribution of loop sizes obtained in this manner
is consistent with the scale invariant distribution (see, e. g., Vachaspati
and Vilenkin, 1984). 

This very exciting experimental development complements work on superfluid 
He$^4$ (where the density of vortex lines is typically much higher, but where
they are also harder to detect). It also opens up the possibility of pressure 
quenches into either of the He$^3$ phases. In particular, phase diagram 
of He$^3$ allowes for a variety of quench trajectories (see Fig.~7), some 
of which raise the possibility of addressing other interesting issues such 
as the creation of topological defects in tuneably weak first order phase 
transformations, or even in a quick succession of the second and first order 
phase transitions.

\noindent
{\bf Quench Into Bulk Superconductors}

\nobreak
Superconductors of type~II have a Landau-Ginzburg form of the free
energy, Eq.~(2.43), and a negative surface energy (so that they can
accommodate vortex lines).  We can therefore repeat the argument
we have put forward for the superfluid phase transition to compute
the density of vortex (flux) lines which obtains as a result of a rapid
quench.  Again, we shall rely on the freeze-out time scale $\hat t$ and
the corresponding correlation length of the order parameter $\hat\xi
= \xi (\hat t)$.  The differences between the case of He$^4$ and
superconductors will be not in the key ideas behind the estimates of
the vortex line density but, rather, in the different experimental
requirements, which will make superconducting bulk quenches more
difficult to carry out.  Moreover, the consequences of the
superconducting quenches may not be easy to investigate (although,
as we shall see in the next section, loop geometry may help!).

As before, we imagine the quench proceeding on a time scale
$\tau_Q$, with the distance away from the phase transition
(measured by the relative temperature $\epsilon$) changing as
$\epsilon (t) \cong t/\tau_Q$, Eq.~(3.2).  As the relative temperature
varies, the relaxation time scale is also changing, $\tau =
\tau_0/|\epsilon|$, so that sufficiently near the phase transition
instant (where $\epsilon = 0$), dynamics of the order parameter
becomes too sluggish to adjust the correlation length $\xi$ to the
equilibrium value given by $\xi_0 \sqrt{|\epsilon|}$, Eq.~(2.48).  The
switch from the adiabatic to the impulse regime will happen (as it was
the case for superfluids) at the freeze-out instant:
$$
\hat t = \sqrt{\tau_0\tau_Q}\;, \eqno (3.30)
$$
where $\tau_0$ is the characteristic time given by Eq.~(2.47).
Consequently, we can evaluate
$$
\hat t \cong 1.225 \sqrt{\tau_Q/T_c}\,[\mu{\rm s}]\;, \eqno (3.31)
$$
where $\tau_Q$ is in seconds and the critical temperature is
expressed in Kelvins.  The frozen-out correlation length is then given
by
$$
\hat\xi = 10^{-2} \left(\xi_0/1000\,\hbox{\AA}\right)
\tau^{1/4}_Q [{\rm cm}] \eqno (3.32)
$$
and corresponds to the critical relative temperature:
$$
\hat\epsilon = \epsilon (\hat t) = 1.225 \cdot
10^{-6}/\sqrt{T_c\tau_Q}\;. \eqno (3.33)
$$
This $\hat\epsilon$ is rather small: Therefore, it might not be
easy to go through the phase transition both sufficiently quickly and
sufficiently uniformly to validate the standard prediction
$$
\ell = 1/\hat\xi^2 = 10^4 (1000\,\hbox{\AA}/\xi_0)^2 \tau^{1/2}_Q
[{\rm cm}^{-2}]\;. \eqno (3.34)
$$
The anticipated initial flux line density is therefore much smaller
than was the case for superfluids.  The decrease in estimated $\ell$
is mainly due to the increase of typical $\xi_0$, which, for a typical
superconductor, is two orders of magnitude bigger than for the
superfluid He$^4$.  One could, of course, increase anticipated initial
$\ell$ by selecting to work with high-temperature superconductors
(which have much smaller $\xi_0 \sim 10$~\AA).  This strategy may
be worth pursuing, although the switch to high-$T_c$ materials is
likely to be accompanied by a specific set of problems (but, possibly,
also by some advantages).  In particular, the Landau-Ginzburg theory
may not be as accurate for the reasons we have explored before.
Moreover, our estimates of $\tau_0$ will not apply, as they are based
on the microscopic theory (BCS) which does not apply to high-$T_c$
superconductors.

Let us now return to the experimental complications.  To begin with,
the pressure quench will not result in as big a change of critical
temperature as is the case for $T_\lambda$ in He$^4$.  There is, of
course, some sensitivity of critical temperature in superconductors
on applied pressure (see, e.g., Lynton, 1963), but the effect is rather
small and may be difficult to exploit experimentally.  Critical
temperature in high-$T_c$ material is more sensitive to pressure
(Neumeier and Zimmerman, 1993; Neumeier, 1994). This 
may be another reason (in addition to smaller $\xi_0$) which
could make them interesting for our purposes.

Temperature quench---rapid cooling of the sample---may be
nevertheless the most practical way to precipitate the phase transition.
Unfortunately, this is a diffusive process.  Therefore, it is likely to
be relatively slow.  Moreover, heat transport is (of course) driven by
temperature gradients.  Thus, limiting the gradient of the relative
temperature to much less than the value set by the ratio:
$$
\hat g = \hat\epsilon/\hat\xi\;, \eqno (3.35)
$$
may not be an easy task.  Gradient $\hat g$  evaluates to:
$$
\hat g \approx 1.225 \cdot 10^{-4} (1000\,\hbox{\AA}/\xi_0)
T^{-1/2}_c \tau^{-3/4}_Q [{\rm cm}^{-1}]\;, \eqno (3.36)
$$
when Gorkov equations (Gorkov, 1959; Tinkham, 1985) apply.  Again,
one may be tempted to consider high-$T_c$ materials, but perhaps a
more practical suggestion would be to work with thin layers of
superconductors.  This 2-D strategy allows one to transport the heat
in the direction perpendicular to the plane in which the
superconducting state will form.  Thus, one may be able to violate the
constraint imposed by Eqs.~(3.35)--(3.36) without invalidating the
reasoning which has led to the estimate of $\hat\xi$, Eq.~(3.32).  This
two-dimensional sample strategy may have one additional
advantage:  The interior of the bulk superconductor is difficult to
probe.  However, flux lines can be detected when they emerge
from the sample.  In a sample with a thickness of the order of $\hat\xi$
($\sim 10^{-2}$~cm) or even somewhat larger, all of the vortex lines
can be ``counted'' where they pierce the surface.  Moreover, using 2-D
geometry may help in slowing down annihilation of vortex lines as one
may be able to supply in a controlled fashion sufficiently many
pinning sites to prevent or at least significantly impede vortex line
migration.

The estimate of the vortex line density is now---in the 2-D
case---given of course by Eq.~(3.24) but with a somewhat revised
interpretation.  We are led to expect a flux line to emerge from
within the area of the order of $\hat\xi$.  Thus, the surface density of
the vortex line endings is given by
$$
\Sigma_i \cong 1/\hat\xi^2\;. \eqno (3.36)
$$

Scattering of neutrons from a sample undergoing a superconducting
phase transition may be a way to peer inside a truly 3-D fragment of
material.  Neutrons have a magnetic moment and will scatter from
the network of flux-filled vortex lines, which should in principle
allow one to deduce vortex line density and perhaps even get
some information about their geometry.

It has also been suggested (Rudaz, Srivastava, and Varma, 1994) that
an electron microscope could be used to image individual flux lines,
although it is harder to imagine how one could use it to track a
network of lines in the course of the phase transition.

So far we have ignored one more obvious complication.
Superconducting vortices are associated with the gauge field ($\vec
B$) which has to be screened in the experiments.  Moreover, the
relative importance of the order parameter and of the gauge field in
the phase transitions with the local gauge symmetry breaking is not
entirely obvious (Rudaz and Srivastava, 1993; Kibble and Vilenkin,
1995).  This last remark may be regarded more as a motivating
factor (rather than as a reason for discouragement):  There are real,
deep, and interesting questions and scenarios of defect formation
in theories with local gauge symmetries.  It is however also
conceivable that transport of the field through the material on the
verge of becoming a superconductor may complicate matters in a
more mundane manner.

\bigskip


\mag=\magstep1
\hsize 6.5truein
\vsize 8.9truein
\lefthyphenmin=3
\clubpenalty 10000
\brokenpenalty 10000
\displaywidowpenalty 10000
\pretolerance 1000

\def\etal{{\it et~al.}}
\def\ul#1{$\underline{\smash{\vphantom{y}\hbox{#1}}}$}
\def\lap{\hbox{~{\lower -2.5pt\hbox{$<$}}\hskip -8pt\raise
-2.5pt\hbox{$\sim$}}}
\def\gap{\hbox{~{\lower -2.5pt\hbox{$>$}}\hskip -8pt\raise
-2.5pt\hbox{$\sim$}}}
\def\degr{^\circ}

\baselineskip 20pt

\noindent
{\bf 4.\quad COSMOLOGICAL EXPERIMENTS IN ANNULAR GEOMETRY}

\nobreak
Let us consider a system undergoing a symmetry-breaking phase
transition.  In the vicinity of the freeze-out instant, $\hat t$, the fate
of its order parameter is sealed:  Phase differences in the
neighboring domains are set.  Hence, the distribution of the
topological defects is also essentially decided.  Each vertex defined
by the domains with the phase differences which add to $\sim\! 2\pi$
will become a section of a topological defect---a superfluid vortex,
a flux line in a type~II superconductor, a disclination
 line in a liquid
crystal, or a cosmic string.  This phase difference can be computed by
adding the phases along a path which circumnavigates the vortex and
which has a radius comparable to the size of the frozen-out domains.

The same postulate of the independent choice of the phase in
independent domains also allows one to compute the total phase
difference along a path with much larger circumference $C$, which
traverses many (${\cal N}$) domains.  The total phase difference
along such a path is given by
$$
\Delta\theta_C \cong \sqrt{{\cal N}} = \sqrt{C/\hat\xi}\;. \eqno (4.1)
$$
Already in the bulk experiments we have discussed so far this
statement is of some interest, as it implies that the vortex lines will
tend to be more anticorrelated than would be the case
if their directions were assigned at random.  This is easy to see.  For,
when the vortex line orientations are selected randomly, the net
circulation along a path defined by a circle of size $r$ (area $\pi
r^2$) would increase with the square root of the number of strings
enclosed, and hence
$$
\Delta\theta \sim \sqrt{\pi r^2/\hat\xi^2}\;. \eqno (4.2)
$$
This implies $\Delta\theta$ increasing with the circumference rather
than with the square root of the circumference [Eq.~(4.1)], as the
more careful analysis implies.  This discrepancy between the two
predictions, Eqs.~(4.1) and~(4.2), can be, of course, readily settled in
favor of Eq.~(4.1), not just by the discussion we have already carried out but 
by an additional observation (Vachaspati and Vilenkin, 1984) that the string 
network generated in the course of a rapid phase transition will consist
($\sim 70$\%) of a long string which will have to cross any surface
spanned by $C$ in opposite directions more often than a random
collection of vectors.  The rest (30\%) of the length is due to (mostly
small) loops which will typically not contribute to $\Delta\theta_C$
unless $C$ passes through their centers (i.e., unless they are "strung"
on $C$ like beads).

The purpose of this section is to discuss how Eq.~(4.1) can be tested
directly in both superfluids and in superconductors.

\noindent
{\bf Superfluid in an Annulus}

\nobreak
Consider quench in an annular container of superfluid He$^4$ (see Fig~8).  The
phase difference locked out in the course of the quench is given by
Eq.~(4.1) with
$$
\hat\xi = \xi_0/|\hat\epsilon|^\nu\;, \eqno (4.3)
$$
where $\hat\epsilon$ is given by $\sqrt{\tau_0\/\tau_Q}$, Eq.~(3.5),
and $\nu = 2/3$ in the renormalization group treatment of He$^4$.
Therefore, in accord with the equation (2.27) which relates the phase
gradient with the velocity of the superfluid, one expects (after the
quench) a flow in a random direction with the velocity of
$$
v = {\hbar \over m} \cdot {1 \over C}\sqrt{C \over \hat\xi} =
{\hbar \over m}\sqrt{1 \over C\hat\xi}\;. \eqno (4.4)
$$

This is obviously an intriguing prediction, especially since it
evaluates to measurable velocities.  To estimate the resulting $v$,
we can further calculate
$$
v = {\hbar\over m}\; {|\hat\epsilon |^{\nu/2} \over \sqrt{\xi_0C}}
\approx 0.8 {\left(\tau_0/\tau_Q\right)^{\nu/4} \over \sqrt{C}}\,
\hbox{[cm/s]}\;. \eqno (4.5)
$$
Above, $\tau_0 \sim 8.5 \cdot 10^{-12}$~s, and $C$ is measured in
centimeters.  Thus, for $\tau_Q$ of the order of milliseconds and $\nu
= 2/3$ one obtains
$$
v \simeq 0.4 \left(\tau_Q \hbox{[$\mu$s]}\right)^{-1/6}\big/\sqrt{C
[{\rm cm}]}\, \hbox{[mm/s]}\;. \eqno (4.6)
$$
This is certainly a macroscopic (if not very large) velocity.
Moreover, Eq.~(4.6) is likely to be an underestimate, since Eq.~(4.1)
contains an implicit assumption that the phase difference between
the domains separated by $\hat\xi$ is approximately a radian, while
the more usual assumption would be to set that
difference at $\pi$ or $2\pi/3$, which would increase
the estimate given by Eq.~(4.6) above 1~mm/s.  A still higher
estimate of the quench-generated velocity (probably of the order of
cm/s) would follow if we used a lower bound on the quench-generated
vorticity implied by the experimental results of the Lancaster group
(Hendry \etal, 1995) to infer the relevant $\hat\xi$.

In any case, locking out a finite ``phase around the loop'' as a result of
a rapid symmetry-breaking phase transformation is the essence of
the cosmological scenario for the generation of topological defects.
Performing the experiment in an annulus is bound to be a challenge,
but the dramatic prediction we have outlined above makes it a
worthwhile effort.  Moreover, the locked-out phase in the loop has one
additional advantage over the vortex creation in the bulk.  The
resulting winding number $n_w$:
$$
n_w = \Delta\theta_C/2\pi \eqno (4.7)
$$
is [to an excellent approximation (Tilley and Tilley, 1986)]
independent of time in the superfluid phase.  Thus, quench can set up
a persistent superflow with a macroscopic, measurable velocity.  This
velocity is directly related to the frozen-out domain size $\hat\xi$
through Eq.~(4.4) and can be---in contrast to the rapidly decaying
vortex line density $\ell$ [see Eqs.~(3.8)--(3.10)]---measured at
leisure long after the quench.

This discussion brings up a few questions.  Let us start with an
apparent paradox:  How can a system which did not rotate
before the phase transition acquire a finite angular momentum
[corresponding to the velocity of $v$, Eqs.~(4.4)--(4.6)] as a result of
an axisymmetric pressure quench?  A facile answer to this question
would be to simply point out that the considerations concerning the
order parameter of the superfluid above do not respect the law of
conservation of angular momentum.  Thus, for example, when an
annular container with superfluid He$^4$ is cooled, the density of the
superfluid is changing; and yet it is known that the phase
difference (and, consequently, the velocity but not the superfluid
angular momentum) remains conserved (Tilley and Tilley, 1986;
Donnelly, 1991).  This is certainly a well-established fact, but it does
not settle the basic issue underlying the angular momentum
generation paradox I have sketched above.  In other words, it does
not say when and to what accuracy the winding number $n_w$ along
some closed path $C$ will be a ``good'' (conserved) quantum number.
For instance, in a bulk superfluid, for a fixed closed $C$, $n_w$ is
certainly not conserved in the course of the evolution of the string
network which follows the quench.  This is because strings can move
across $C$, a process which obviously alters $n_w$.

By contrast, in an annulus, cooling (or heating) of the superfluid will
lead to a change of the superfluid angular momentum (as a result of a
change of the Bose condensate density $\rho_s \sim |\Psi|^2$) but
the global conservation law will be nevertheless respected, since
the superfluid can ``push off'' the normal fluid, which will in turn
dissipate the excess momentum on the walls of the container.

The first obvious remark in the wake of the above discussion is to
note that, in order to be safe from the ``bulk'' decay of the vortex
line network, one must choose the small radius of the annulus to be
of the order of $\hat\xi$.  Otherwise, the resulting ``phase around the
loop'' will be modified by the vortex lines enclosed within the torus.
Thus, one will be forced to work with capillaries, as the typical values
of $\hat\xi$ are close to $\sim 10^4$~\AA.  This narrow annulus may
suffer from the possibility of thermally activated transitions
(especially near $T_c$), and this process may cause the decay of
$n_w$.  Perhaps the most likely mechanism for such slow decay of the
superflow would be due to creation of vortex
lines, which can then migrate across the annulus, thus changing $n_w$
by one unit.

The coincidence of the size of the small diameter of the annulus with
the frozen-out correlation length $\hat\xi$ contains a crucial clue to
the resolution of the angular momentum generation paradox I have
described at the beginning of this discussion.  This is because we can
now estimate the angular momentum involved in the superflow as
$$
J_s = \left(\rho_s \cdot {1 \over 4}\pi\hat\xi^2 \cdot C\right)
\cdot v \cdot r\;, \eqno (4.8)
$$
where $r = C/2\pi$.  The only conceivable source for angular
momentum $J_s$ (which, for a typical superfluid $\rho_s \sim
0.1$~g/cm$^3$,
$\hat\xi = 10^4$~\AA, and $C = 1$~cm is, in accord with
Eqs.~(4.4)--(4.6), $J_s
\sim 5 \cdot 10^{-11}$~g~cm$^2$/s) would be the angular momentum
of Brownian motion, which is given by
$$
J_T = \left(k_B T \cdot M \cdot r^2\right)^{1/2}\;. \eqno (4.9)
$$
Here $M$ is the mass of the involved material [which is given by $M =
\rho_s \cdot \pi\hat\xi^2 \cdot C/4$ in Eq.~(4.8)].  
The above equationcan be readily derived when the kinetic energy of
rotation (given by $J_T^2/(2 M r^2)$) is equated to the thermal $k_B T / 2$
per degree of freedom.
At $T = T_\lambda
= 2\degr$K for the same $\rho_s$ and $C$ we can estimate typical
$J_T \sim 10^{-13}$~g~cm$^2$/s, which is much less (several orders
of magnitude) than $J_s$.  Thus, at the first sight, it would seem
unlikely that $J_T$ could be the origin of $J_s$.

However, the equation
$$
J_s = J_T \eqno (4.10)
$$
would have to be satisfied not deep in the superfluid phase, where
$\epsilon \sim 1$ and $\rho_s \sim 0.1$~[g/cm$^3$], as we have
assumed above, but, rather, at the freeze-out instant $\hat t$, for the
corresponding relative temperature $\hat\epsilon \ll 1$.  Now, it is
well known that
$$
\rho_s = \rho_s (0) \epsilon^\nu\;, \eqno (4.11)
$$
where $\nu = 2/3$ in the renormalization group theory and in the
experiments (Ahlers, 1976), and $\rho_0 \simeq 0.36$~g/cm$^3$.  In
accord with Eqs.~(4.8)--(4.10) we get
$$
\rho_s \left(\pi \hat\xi\right) = 4 \left(m/\hbar\right)^2 k_B
T_\lambda\;, \eqno (4.12)
$$
which is to be satisfied at $\hat\epsilon$.  But the left-hand side of
Eq.~(4.12) is {\it independent\/} of $\hat\epsilon$ as $\hat\xi =
\xi_0\epsilon^{-\nu}$.  Hence, we are led to an unexpected
temperature-independent relation:
$$
\rho_s (0) \xi_0 = 4{m^2 \over \hbar^2}\cdot k_BT_\lambda/\pi
\eqno (4.13)
$$
between the density of liquid He$^4$, its correlation length (both at
absolute zero), mass of the helium atom, fundamental physical
constants $\hbar$ and $k_B$, and the critical temperature
$T_\lambda$.  The left-hand side of Eq.~(4.13), for the data cited by
Ahlers (1976) is
$$
\rho_s (0) \xi_0 \cong 1.44 \cdot 10^{-8} \hbox{[g cm$^{-2}$]}\;. \eqno
(4.14)
$$
The right-hand side is
$$
4\left(m^2/\hbar^2\right)k_B T/\pi \simeq 1.4 \cdot 10^{-8} \hbox{[g
cm$^{-2}$]}\;. \eqno (4.15)
$$
This coincidence is striking and inspires confidence in the Brownian
motion explanation of the origin of the angular momentum locked out
by the quench.  Indeed, Eq.~(4.13) has been postulated before in the
discussion of healing length in superfluid He$^4$ (Ahlers, 1976; Ferrel
\etal, 1969).

\noindent
{\bf Quench in a Superconducting Loop}

\nobreak
From the experimental point of view, trapping of the winding number
$n_w$ by a rapid quench in a loop of a superconducting wire may be
simpler than the corresponding experiment in the superfluid He$^4$.
In particular, the detection of the effect---which, in a superfluid
He$^4$ involves measurement of small amounts of angular
momentum---should be now much simpler, as it suffices to measure
the magnetic field associated with the number of flux quanta
trapped during the quench.  These first impressions may be
indeed valid as far as the experimental situation is concerned; but
from the point of view of relating such experiments to cosmology,
they are, I believe, too optimistic.  As we shall see below, quench in a
superconducting loop is complicated by the very presence of the
magnetic (gauge) field, which makes the theoretical analysis more
difficult and which will also complicate experiments.

We start by, in effect, repeating the ``generic'' prediction based on
the cosmological scenario for the number of flux quanta generated in
converse of a rapid quench:
$$
n_\Phi = n_w = {\Delta\theta_C \over 2\pi} = (2\pi)^{-1}
\sqrt{C/\hat\xi}\;. \eqno (4.16)
$$
Here $n_\Phi$ stands for the number of trapped flux quanta.  So far,
the effect of the gauge fields is ignored.  For a loop of radius $r =
1$~cm and a frozen-out correlation length $\hat\xi \sim 10^{-2}$~cm
[see Eq.~(3.22)] this yields $n_\Phi \simeq 3$.  The corresponding flux
would be small but easily measurable by the techniques utilizing
SQUID's (Tinkham, 1985).

To arrive at the above prediction, Eq.~(4.16), we have paid attention
solely to the order parameter.  But the evolution of the magnetic
field trapped by the quench may play a role in the final outcome.  To
consider this possibility, let us first note that the energy associated
with the trapped flux $\Phi$ is nonnegligible:
$$
E_\Phi = \Phi^2/2L = n^2_\Phi \cdot \Phi^2_0/2L = n^2_\Phi E_0\;.
\eqno (4.17)
$$
Above, $E_0$ stands for the energy associated with a single flux
quantum $\Phi_0 = hc/2e$ [Eq.~(2.52)] trapped in a loop with the
inductance $L$:
$$
E_0 = \Phi^2_0/2L\;. \eqno (4.18)
$$
In the circular loop of wire the inductance (measured in Henrys) is given by
$$
L \cong 4\pi \cdot 10^{-9} r\,\ln (r/a)\; [H]\;, \eqno (4.19)
$$
where $r$ is the radius given in centimeters and $2a$ is the
diameter of the wire.  This results in typical energies:
$$
E_0 \sim 2.5 \cdot 10^{-16} r^{-1}\; [{\rm erg}]\;, \eqno (4.20)
$$
where we have set $r/a = 1000$.  Energy of thermal excitations
is---for comparison---given by
$$
E_T \simeq {1 \over 2} k_B T_c \simeq 7 \times 10^{-17} \cdot T_c\;
[{\rm erg}] \eqno (4.21)
$$
in the vicinity of the superconducting phase transition temperature
$T_c$ (which is given in $\degr$K).

This comparison---$E_\Phi$ with $E_T$---is relevant because we
would like to be able to trace the origin of the trapped quanta to
either a ``cosmological'' freeze out of the order parameter or to the
trapping of the thermal or other fluctuations of the electromagnetic
field.  This second possibility is obviously important, as it is well
known that in the course of a slow phase transition in an external
field the flux trapped by a superconducting loop is determined by
that magnetic field.  The above considerations allow us to at least
differentiate between the experimentally testable consequences of
the two alternatives.  Predictions based on the ``cosmological''
Eq.~(4.16) depend only on the quench rate (which determines
$\hat\xi$) and on the circumference $C$ of the loop.  By contrast,
trapping of the thermal fluctuations of the field will be affected by
the self-inductance $L$, which can be changed without altering $C$
(by, for example, coiling up the loop into a solenoid).  Thus, if the
trapped flux has its origin in thermal fluctuations of the trapped flux,
the expected value of the flux will be given by
$$
{\delta\Phi^2_T \over 2L} = {1 \over 2} k_B T\;. \eqno (4.22)
$$
By contrast, freeze out of the order parameter leads to the
prediction:
$$
\left(\delta\Phi/\Phi_0\right)^2 = n^2_\Phi = C/\hat\xi\;. \eqno
(4.23)
$$
The number of trapped quanta should also slowly scale with the
quench rate, since $n_\Phi$ in the Landau-Ginzburg theory varies as
$$
n_\Phi \sim \tau^{-1/8}_Q\;. \eqno (4.24)
$$

The experiment should then be able to tell which of the two
alternatives---the one driven by the fluctuations of the order
parameter or the one in which the gauge field dominates---is
realized ``in practice.''  This question is of cosmological interest
since, as we have already indicated, the opinion concerning
overwhelming importance of the order parameter in the creation of
topological defects is not unanimous (Rudaz and Srivastava, 1993;
Kibble and Vilenkin, 1995).  However, it should be noted that there are
significant differences between the physics which is relevant in the
cosmological (bulk) context and the one which will play an important
role in the superconducting loop experiment considered here.  In
particular, the gauge field in the bulk phase transition acquires
mass---and, therefore, becomes limited in its range to the
penetration depth
$\lambda = \kappa\xi$---as the phase transition is taking place.
Thus, the range over which the order parameter may become causally
connected will be in any case limited by an exponential to a finite
distance, even if the phase information could propagate with the
gauge field.  By contrast, quench in a loop leaves the magnetic field
inside the loop massless.  Therefore, if gauge fields were to play a
dominant role in setting up relative phases of the broken symmetry
Bose condensate, the infinite range and the speed with which
different sections of the loop can influence one another could prove
significant.

Detailed consequences of this competition between the order
parameter (Bose condensate) and the gauge (magnetic) field are
difficult to predict in general.  We shall explore some of the relevant
physical phenomena below.  We start by noting that, in addition to
the relaxation time $\tau$ and the quench time scale $\tau_Q$ (which
in turn decides the freeze-out instant $\hat t$), there is at least one
more time scale which limits the speed with which changes in the
magnetic field (or Bose condensate) configuration can take place.
The loop is an $R$-$L$ circuit, so the flux it encloses can change no
faster than on a time scale:
$$
\tau_{RL} = L/R\;. \eqno (4.25)
$$
Here $R$ is the resistance.  In the vicinity of the superconducting
phase transition, this time scale is rapidly increasing, as $R$ tends to
zero.  Moreover, it is dominated by the few sections of the loop
which are still normal, because of, for example, slightly different
critical temperature $T_c$ in different sections of the wire.

This situation (and more generally, any configuration of a
two-dimensional $\Psi$ in an approximately 1-D space) can be
represented in a diagram shown in Fig.~9.  The
quantity of interest in our considerations---known as the {\it
fluxoid\/} in the superconductor physics [see Eq.~(2.55)]---can be
used to define the gauge-invariant winding number:
$$
n_w = \oint_C \left(\vec\nabla\theta - {2\pi \over \Phi_0}
\vec A\right) \vec{dl}\;. \eqno (4.26)
$$
In effect $n_w$ is the number of times the phase of the order
parameter wraps around along the loop $C$.

In equilibrium, the amplitude of $\Psi$ is set by $\sigma$, Eq.~(2.33).
The turns are approximately equally spaced and can be thought of as regular
(although their distribution is gauge-dependent).  As the phase
transition temperature is approached from below, the number of
turns is constant, but $\sigma \sim \sqrt{|\epsilon|}$ becomes smaller,
which may eventually allow for changes in $n_w$.  We shall consider
processes which can cause changes of $n_w$ below.  Let us begin with
a situation where the loop $C$ is superconducting except for a small
section somewhere along its circumference.  Then the winding number
will not be fixed.  

When a system exists in this state for a time long compared to the
time scale on which the flux through the loop can change, the
configurations of $\Psi$ will explore distribution determined, on the
one hand,  by the energy of these configurations (set both by $\Psi
(x)$ and by the magnetic field associated with it) and by the external
noise sources which can pump energy into the system both through
the magnetic field and through $\Psi$.  When the temperature in the
normal section of the wire is lowered (so that the two ends of $\Psi
(x)$ are joined together), the winding number cannot change any
more.

An experiment which explored this regime (although with a somewhat different 
interpretation in mind) has been actually already carried out.  Tate, Cabera, 
and Felch (1984) have heated up a small section of a superconducting
niobium loop with a laser.  After a laser was turned off, the number of
trapped flux quanta inside the loop was measured.  The random
number of quanta had a dispersion corresponding to the temperature
6.78$\degr$K, which is less than the critical temperature of pure
niobium ($T_c
\cong 9.17\degr$K).  They have interpreted this as a result of thermal
excitations with the dispersion set by the local critical temperature
in the heated-up section of the loop.  The discrepancy in the value of
$T_c$ was explained as a result of local fluctuation of the critical
temperature in the loop.  In view of our preceding discussion, one
might venture an alternative explanation. The lower temperature associated
with the dispersion of the number of the locked out quanta may be 
a result of the coupling of the system (including the
order parameter) with both the ``hot spot'' illuminated by the laser
and with the heat bath (which was presumably at a still lower
temperature than the reported 6.78$\degr$K).

\noindent
{\bf Activation Processes in a Superconducting Loop}

\nobreak
The above discussion has prepared us to consider the problem of the
phase transition in a loop from the new vantage point with the
intuitive understanding based on Fig.~8.  The first conclusion is
already at hand.  When the phase transition occurs so nonuniformly
that the last ``break'' in the continuity of the Bose condensate
disappears only well after the time scale needed for equilibration 
of the rest of the loop ($t>\tau_{RL}$) we can expect a distribution set
(or, at least, significantly influenced) by the thermal and other
fluctuations driving the system.  One of these configurations
becomes then topologically ``trapped'' when the whole loop becomes
superconducting.

By contrast, when the phase transition occurs simultaneously along
the loop, we recover the picture we have discussed in the
cosmological context.  Different sections of the loop reach the
broken symmetry phase independently.  Therefore, at the end of the
quench, the flux trapped inside $C$ will be set by the winding number
determined by Eqs.~(4.1) and~(4.16).

It should be noted that the value of the flux is not
constrained to the multiple of the number of flux quanta but, rather,
by the fluxoid quantization condition, Eqs.~(2.55) and~(4.26).  The two
are equivalent only in the limit where the penetration depth
$\lambda$ is small compared with the thickness of the wire, $\lambda
\ll 2a$.  When this is not the case (i.e., when $\lambda > a$), the flux
inside $C$ may be quantized in units smaller than $\Phi_0$.  This is
because in that limit both the velocity of the Cooper pairs and the
vector potential are important (Blatt, 1961; Bardeen, 1961).  This
can be investigated in the case where the current is approximately
independent of the location inside the wire (which is the case when
$\lambda > 2a$).  The velocity of the Cooper pairs is then given by
the gradient of the phase which is not compensated by the magnetic
flux:
$$
v = \hbar q/m^\ast = \hbar q/2m\;. \eqno (4.27)
$$
Here, $m^\ast$ is the mass of a Cooper pair and $q$ is the residual
wave vector:
$$
q = {2\pi \over C} \left(n_\Phi - {\Phi \over \Phi_0}\right)\;. \eqno
(4.28)
$$
Above, the integer $n_\Phi$ is chosen so that $qC$ is in the interval
($-\pi$, $\pi$).

Equilibrium values of the magnetic flux $\Phi$ through $C$ will be
determined by the minima of the total energy associated with a
certain flux and with the kinetic energy corresponding to the
velocity of the charge carriers, Eqs.~(4.27) and~(4.28):
$$
E = \Phi^2/2L + N_C m^\ast v^2/2\;. \eqno (4.29)
$$
Here $L$ is the inductance and $N_C$ is the total number of Cooper
pairs in the loop $C$.  It can be obtained from the
(temperature-dependent) density of the Cooper pairs $n^\ast_C =
|\Psi|^2$:
$$
N_C = C\pi a^2 \cdot |\Psi|^2\;. \eqno (4.30)
$$
Using the above equations, it is straightforward to show that $E$ is
minimized when the flux is given by
$$
\Phi = \Phi_0 \cdot n_\Phi/(1 + E_0/E_K)\;. \eqno (4.31)
$$
Here, $E_0$ is defined by Eq.~(4.18) and corresponds to a single
trapped quantum of flux, while $E_K$ is the kinetic energy of the
charge carriers for the velocity corresponding to the gradient of
phase given by $2\pi/C$:
$$
E_K = {1 \over 2} N_C m^\ast \left({\hbar \over m^\ast} \cdot
{2\pi \over C}\right)^2\;. \eqno (4.32)
$$
It follows that, in equilibrium, the flux in a loop of a given
self-inductance $L$ and circumference $C$ will be quantized in units
of
$$
\tilde\Phi_0 = \Phi_0/(1 + E_0/E_K)\;. \eqno (4.33)
$$
Furthermore, in the vicinity of $T_c$ the winding number $n_w ~=~ n_{\Phi}$ can
be altered by thermal activation.  The potential barrier which
separates different integer values of the winding number is
approximately given by
$$
F_B \cong E_K/2\;, \eqno (4.34)
$$
for the small values of $n_w$.  This energy can be easily small enough
to be of the order of $k_BT_c$.  Thus, in the course of the phase
transition and already below $T_c$, $n_w$ can still change and may be
driven towards a value set by thermal (or other) fluctuations which
couple to the system.

The ``unwinding'' transition we have described above occurs
along all of the circumference $C$.  By contrast, one can imagine a
localized thermal excitation involving a correlation-length sized
section of the superconducting wire.  The free energy barrier
associated with such a transition is approximately given by the
volume in which it takes place times the specific free energy of the
symmetric state.  The exact expression (see Chapter~7 of Tinkham, 1985)
turns out to be
$$
\Delta F = {4\sqrt{2} \over 3} \cdot {\alpha^2 \over \beta} \cdot
A\xi\;, \eqno (4.35)
$$
where $A$ is the cross section of the wire.  Typically $\Delta F$
[Eq.~(4.35)] is significantly larger than $F_B$, Eq.~(4.34).  However,
the time scale associated with the thermal excitations within one
correlation length (given by $\tau$) is likely to be much smaller than
the time (probably of the order of $\tau_{RL}$) which will set the
``whole loop'' rate of transitions guarded against by the barrier
$F_B$, Eq.~(4.34), we have discussed before.  Thus, it is not clear
which of these two processes will dominate or whether any of them
will have a sufficient rate to reset the number of trapped quanta
from the value determined by the frozen-out fluctuations of the
order parameter, Eq.~(4.23), to the estimate of Eq.~(4.22), set by
thermal fluctuations of the flux.

Last but not least, we note that the study of the dynamics of the
order parameter in a superconducting loop with one (or more) ``gaps''
is an intriguing subject in its own right.  Thus, one could create a
number of independent ``domains'' in the superconductor by cutting
up the order parameter (i.e., by heating up the loop locally).  These
superconducting domains can be then welded together, and the
resulting trapped winding number (or the resulting flux) compared
with theoretical expectations, which would have to be based on some
compromise between the order parameter freeze out and thermal
fluctuations driven activation process we have discussed above.  One
can imagine designing experiments in which the individual domains
are small enough so that $n_w$ is given by Eq.~(4.16), or alternatively,
where the flux is essentially thermal, with the expectation value set
by Eq.~(4.22).  It should be perhaps pointed out that such
experiments could also shed a new light on the fascinating questions
about the nature of the phase of the wave function of the Bose
condensate raised by Anderson (1986) and Leggett (1980) and
intimately related to the interpretational issues of quantum theory
(Wheeler and Zurek, 1983; Zurek, 1991).

\bigskip


\mag=\magstep1
\hsize 6.5truein
\vsize 8.9truein
\lefthyphenmin=3
\clubpenalty 10000
\brokenpenalty 10000
\displaywidowpenalty 10000
\pretolerance 1000

\def\etal{{\it et~al.}}
\def\ul#1{$\underline{\smash{\vphantom{y}\hbox{#1}}}$}
\def\lap{\hbox{~{\lower -2.5pt\hbox{$<$}}\hskip -8pt\raise
-2.5pt\hbox{$\sim$}}}
\def\gap{\hbox{~{\lower -2.5pt\hbox{$>$}}\hskip -8pt\raise
-2.5pt\hbox{$\sim$}}}
\def\degr{^\circ}

\baselineskip 20pt

\noindent
{\bf 5.\quad COSMOLOGICAL IMPLICATIONS}


\nobreak
The main difference between the dynamics of the phase
transition in the condensed matter and cosmological contexts
arises from a different scaling of the relaxation time scale
with the relative temperature $\epsilon$.  Order parameters
(fields) which are relevant for cosmology obey the equation of
motion with the second time derivative on the left-hand side.
Consequently, relaxation time is
$$
\tau = {1 \over \sqrt{|\alpha|}} \sim {1 \over \sqrt{|\epsilon|}}\;.
\eqno (5.1)
$$
(Above, we have returned to $\hbar = c= k_B = 1$.)  One
immediate consequence of this scaling is that the velocity
with which perturbations of the order parameter can
propagate, given by $\xi/\tau$, remains finite even at $T_c$,
since
$$
\xi = {1 \over \sqrt{|\alpha|}}  \sim {1 \over \sqrt{|\epsilon|}}\;.
\eqno (5.2)
$$
For the purpose of evaluating the initial density of the
topological defects in the cosmological phase
transitions, we should also note that $\alpha' = \alpha/\epsilon$
is approximately
$$
\alpha' \approx \beta T^2_c \eqno (5.3)
$$
in field-theoretic models.

In the early Universe the equation of state is dominated by
radiation.  Thus, the rate at which the phase transition happens
will be determined by the relation
$$
T^2 t = \Gamma M_{PL}\;, \eqno (5.4)
$$
which holds in the radiation-dominated Universe.  Above $t$ is
the time since the ``Big Bang,'' and $M_{PL}$ is the Planck mass.
The coefficient $\Gamma$ depends on the effective number of
different spin states $s$ of relativistic particles:
$$
\Gamma = {1 \over 4\pi} \cdot \sqrt{{45\over \pi s}}\;. \eqno
(5.5)
$$
The quench time scale can be then obtained directly from
Eq.~(5.4) as
$$
\tau_Q = 1/\dot\epsilon = 2\Gamma M_{PL}/T^2_c\;. \eqno (5.6)
$$
Moreover, the relaxation time can be expressed as
$$
\tau = \tau_0\Big/\sqrt{|\epsilon|}\;, \eqno (5.7)
$$
where $\tau_0$ can be evaluated with the help of Eqs.~(5.1)
and~(5.3);
$$
\tau_0 \sim \left(\sqrt{\beta}\, T_c\right)^{-1}\;. \eqno (5.8)
$$
We have all of the ingredients required to compute the
cosmological freeze-out time $\hat t$, which now obeys the
equation
$$
\tau_0\Big/\sqrt{\hat t/\tau_Q} = \hat t\;. \eqno (5.9)
$$
This is the relativistic version of Eq.~(3.3) we have solved
before for superfluids and superconductors, but now the
solution reads:
$$
\hat t = \tau^{2/3}_0 \tau^{1/3}_Q\;. \eqno (5.10)
$$
The difference with the previously obtained $\hat t =
\sqrt{\tau_0\tau_Q}$, Eq.~(3.4), stems from the different
scaling of $\tau$ with $\epsilon$, Eq.~(5.1).

It is now straightforward to evaluate
$$
\hat t = \left(2\Gamma M_{PL}/T_c\right)^{1/3}\Big/
\left(\beta^{1/3}T_c\right) \eqno (5.11)
$$
and to obtain the corresponding $\hat\epsilon$;
$$
\hat\epsilon = \beta^{-1/3}
\left({T_c \over 2\Gamma M_{PL}}\right)^{2/3} \eqno (5.12)
$$
We predict that the freeze out will take place at
$\hat\epsilon$ and that it will determine the overall structure of
the initial configuration of cosmological defects.

The characteristic correlation length would be then of the order
of
$$
\hat\xi = \left({2\Gamma M_{PL} \over T_c}\right)^{1/3} \cdot
{1 \over \beta^{1/3}T_c}\;. \eqno (5.13)
$$
The corresponding density of topological defects obtains from
the usual argument (i.e., one monopole/one $\hat\xi$ section of
a string/one $\hat\xi^2$ section of membrane per volume of
$\hat\xi^3$).  This value of the freeze-out relative temperature
$\hat\epsilon$ is, of course, quite different from what obtains
when the Ginzburg condition is employed.  In that latter case:
$$
\epsilon_G \sim \beta\;. \eqno (5.14)
$$
This would in turn lead to the correlation length of
$$
\xi_G = \left(\beta T_c\right)^{-1}\;, \eqno (5.15)
$$
which is, again, quite different from the estimate of
Eq.~(5.13).  Indeed, the ratio
$$
\hat\xi/\xi_0 = \beta^{2/3} \cdot \left({2\Gamma M_{PL}
\over T_c}\right)^{1/3} \eqno (5.16)
$$
shows that, for the usually assumed small $\beta$, $\hat\xi$
could be smaller than $\xi_G$ for very high temperature phase
transitions.  However, when $T_c  \ll M_{PL}$---that is, for late
phase transitions---$\hat\xi$ is likely to be much larger than
$\xi_G$.

Whether the difference between the resulting initial densities
of topological defects will be of great significance for a
cosmological model is, of course, a separate question.  For
example, in the case of cosmic strings most of the observable
consequences will be determined by the structure of the string
network at late times and will be decided by the dynamics of
string interactions, etc., rather than by the details of their
initial configuration.  On the other hand, when---as is
sometimes proposed---topological defects are used as a
catalyst in baryogenesis, their density will leave an immediate
imprint on the Universe and will be critically important.

\bigskip
\noindent
{\bf 6.\quad DISCUSSION}

\nobreak
We have considered creation of topological defects in course of the 
second order phase transition involving non-conserved order parameter.
This selection was dictated by the importance of such systems for cosmology
as well as by the recent experiment in superfluid He$^4$ which afforded
a laboratory test of the cosmological scenario. The original cosmological 
motivation notwithstanding, defect-forming dynamics of the second order phase
transitions is of enormous interest in its own right.

The key question we have attempted to address concerned the initial density
of the topological defects. This quantity is a witness to the symmetry
breaking dynamics in course of the phase transformation. Two paradigms were put
forward to compute initial density of topological defects. The older one
appeals to the activation process and emphasizes the relevance of the Ginzburg
temperature---the temperature below which thermally activated transitions
become prohibitively expensive---on the initial density of the defects.
This line of reasoning would lead one predict a copious production of defects
anywhere between the Ginzburg temperature $T_G$ and the phase transition 
temperature $T_c$. Moreover, any rapid quench trajectory which  starts within
that region would be expected to lead to more or less the same density of
defects (set by the correlation length at the Ginzburg temperature). 

In contrast to the thermal activation mechanism sketched out above, one can 
consider a scenario in which the density of defects is determined
by the dynamics of the order parameter in the immediate vicinity of the 
critical temperature $T_c$. The characteristic scale (which sets the density
of the topological defects) is decided at the instant when the critical
slowing down makes the order parameter so sluggish that it can no longer keep
up with the changes of the thermodynamic conditions induced by the quench. 
This happens at the relative temperature $\hat \epsilon$ [see Eqs. (3.5)
and (5.12) for the typical condensed matter and cosmological cases, 
respectively].

Freezeout scenario is guaranteed to set the scale of the initial distribution
of cosmological defects when $\hat \epsilon$ is below the relative Ginzburg
temperature $\epsilon_G$. This is because;
$$
|\hat \epsilon| > |\epsilon_G| \eqno(6.1)
$$
implies that the dynamics of the
order parameter remains effectively frozen until $\hat \epsilon$ is reached, 
so there is simply no time for the activation process. In the laboratory phase
transition this condition will in principle depend on the quench rate, but
is in any case likely to be satisfied in the superconductors,
and violated in the superfluid He$^4$. In the cosmological phase 
transitions it translates into a relatively simple inequality:
$$ 
\beta^2 \gap {T_c \over {2 \Gamma M_{PL}}}\;. \eqno (6.2)
$$
This condition follows from Eqs.~(5.12), (5.14), and a demand that freezeout 
should persist as the Ginzburg temperature is being traversed during 
the quench. Thus there is certainly a range where
the freezeout preempts activation ``by default''.

It may seem more surprising that---as it was anticipated in 
the original proposal for ``cosmology in the laboratory'' 
(Zurek, 1984, 1985, 1993), and as it is apparently 
borne out by the experimental results (Hendry et al., 1994, 1995)---the 
freezeout scenario is valid even when inequality (6.1) is grossly violated,
so that the activation events are expected to occur frequently. Apparently,
in the superfluid (and, presumably, also in the cosmological phase transitions)
the larger scale set by the critical slowing down is ``remembered'', and its
memory is never erased by the activation occuring on the smaller scales.

While this may have been a surprise, it was certainly not totally unexpected.
For one thing, the energetic cost of creating large scale fluctuations
is prohibitive. And only perturbations which have sizes comparable with the
characteristic scale of the string laid down during the freezeout can 
appreciably alter initial density of the string network. Moreover, the bulk
($\sim 70\%$) of network is expected to belong to a single long string.
Its geometry may be effected by activation events on the smaller scale 
corresponding to the correlation length at the Ginzburg temperature, but the 
very fact that the phase transition has already taken place implies that 
such small scale fluctuations do not appreciably re-arrange long range order.

Seen in this light, the apparent lack of importance of the Ginzburg temperature
for the initial density of cosmological defects does not appear too surprising.
The symmetry is broken---long range order sets in---at the critical temperature.
Topological defects are, after all, imperfections of that long range order. 
Hence, persistence of the large scale structure imposed at the time when long 
range order comes into being for the first time can be reconciled with the small
scale perturbations caused by thermal activation. 

This is not to imply that the above arguments and the apparent accord between 
the experiment and the scenario which appeals to the freezeout 
(and an even more obvious discord with the consequences of 
the activation scenario) settles all the issues which are worth settling.
On the contrary, understanding of the dynamics of the process emergence of the long range
order in the course of the second order phase transitions is only at its 
inception. Issues which have not been adressed in detail (such as the 
interplay between the freezeout and activation) can be perhaps tackled
by numerical simulation (preliminary results are already in; Laguna and
Zurek, in preparation). They also include mode dependence of the critical
dynamics (Gill and Rivers, in preparation), the influence 
of the gauge fields, as well as a study of problems
specifically designed to address experiments. Moreover, experiments themselves
should not be restricted to just quenches through the critical region. Quenches
from just below the critical temperature, as well as transitions from
the broken symmetry phase to just above $T_c$ followed by a quick return to the
broken symmetry phase should allow one to probe the timescale on which
the memory of the broken symmetry is erased. One can also imagine performing
quenches into the superfluid starting from within the solid phase of He$^4$
or He$^3$. The order parameter responsible for the existence of topological
defects (presumably) disappears when superfluid turns into solid. However, 
second sound in the solid He$^4$ exists, which suggests that at least some 
aspect of the superfluid behavior may persist in the solid as well. It is of 
course difficult to imagine existence of vortices in the solid. But this
(and other similar ``out on the limb'' speculations) can be experimentally
addressed by checking that the topological defects can be created in course 
of the solid-superfluid pressure quench, and by finding out whether 
the topological defects imprinted on the superfluid can weather the transition 
into the solid phase. (If they did -- which seems unlikely -- then it should 
be possible to freeze out and then to ``defrost'' vortex lines.) 

This paper was more a preview of the exciting possibilities rather than 
a review in the traditional sense. Its success should be measured by how much 
it contributes to its own obsolescence. For, it is hoped that the research 
carried out in the exciting areas outlined above will be rapid enough to
surpass and eclipse the developments described here.

\bigskip
\noindent{\bf Acknowledgments}

This research was initiated during a programme ``Formation and Interactions of
Topological Defects'' which was held at the Newton Institute in the Fall of 
1994. I have benefited from discussions with Andreas Albrecht, Alasdair Gill,
Tom Kibble, Peter McClintock, Pablo Laguna, Ray Rivers, and Alex Vilenkin.

\bigskip

\mag=\magstep1
\hsize 6.5truein
\vsize 8.9truein
\lefthyphenmin=3
\clubpenalty 10000
\brokenpenalty 10000
\displaywidowpenalty 10000
\pretolerance 1000

\def\refer{
\par
\parindent=0pt
\hangafter=1
\hangindent=9pt
}
\def\etal{{\it et~al.}}
\def\ul#1{$\underline{\smash{\vphantom{y}\hbox{#1}}}$}
\def\lap{\hbox{~{\lower -2.5pt\hbox{$<$}}\hskip -8pt\raise
-2.5pt\hbox{$\sim$}}}
\def\gap{\hbox{~{\lower -2.5pt\hbox{$>$}}\hskip -8pt\raise
-2.5pt\hbox{$\sim$}}}
\def\degr{^\circ}

\bigskip
\baselineskip 15pt
\noindent
{\bf References}

\nobreak
\refer{}Ahlers, G. 1976, Chap.~2 in The Physics of Liquid and Solid
Helium, ed.\  K.~H. Bennemann \& J.~B. Ketterson (Wiley,
New York), Vol.~2

\refer{}Aitchison, I.~J.~R., \& Hey, A.~J.~G.  1982, Gauge Theories
in Particle Physics (Hilger, Bristol)

\refer{}Albrecht, A., \& Stebbins, A. 1992, Phys.\ Rev.\ Lett., 68,
2121; {\it ibid.} 1992, 69, 2615

\refer{}Albrecht, A., \& Steinhardt, P.~J. 1982, Phys.\ Rev.\ Lett.,
48, 1220

\refer{}Anderson, P. W. 1986, in The Lesson of Quantum
Theory, ed.\ J. deBoer,  E. Dal,  \& O.~Ulfbech,  (Proc.\ Niels Bohr
Centennial Symposium, Copenhagen, October 1985) (North Holland,
Amsterdam)

\refer{}Bardeen, J. 1961, Phys.\ Rev.\ Lett., 7, 162

\refer{}Bardeen, J., Cooper, L.~N., \& Schrieffer, J.~R.  1957, Phys.\
Rev., 108, 1175

\refer{}Bernstein, J., \& Dodelson, S. 1991, Phys.\ Rev.\ Lett.,
66, 683

\refer{}Blatt, J.~M. 1961, Phys.\ Rev.\ Lett., 7, 82

\refer{}Bowick, M.~J., Chandar, L., Schiff, E.~A., \& Srivastava, A.~M.
1994, Science, 263

\refer{}Bray, A.~J.  1995, Advances in Physics, in press

\refer{}Chuang,~I., D\"urrer, R., Turok, N., \& Yurke, B. 1991, Science,
251, 1336

\refer{}de Gennes, P.~G., The Physics of Liquid Crystals (Clarendon Press,
Oxford, 1974)

\refer{}Donnelly, R.~J.  1991, Quantized Vortices in Helium~II
(Cambridge University Press, Cambridge)

\refer{}Ferrel, R.~A., Menyhard, N., Schmidt, H., Schwabl, F., \&
Szepfalusy, P. 1969, Ann.\ Phys., 177, 352

\refer{}Feynman, R.~P. 1954, Phys.\ Rev., 94, 262

\refer{}Gill, A.~J., \& Rivers, R.~J., in preparation

\refer{}Ginzburg, V.~L.,  \& Pitaevskii, L.~P.  1958, Zh.\ Eksp.\
Teor.\ Fiz., 34, 1240; 1958, Soviet Physics---JETP,
34, 858

\refer{}Goldenfeld, N. 1992,  Lectures on Phase Transitions and
the Renormalization Group (Addison-Wesley, Reading,
Massachusetts)

\refer{}Gorkov, L.~P.  1959, Zh.\ Eksp.\ Teor.\ Fiz., 36, 1918;
1959, Soviet Physics---JETP, 36, 1364

\refer{}Guth, A.~H. 1981, Phys.\ Rev.~D, 23, 347

\refer{}Hendry, P.~C., Lawson, N.~S., Lee, R.~A.~M.,
McClintock, P.~V.~E., \& Williams, C.~H.~D. 1994, Nature, 368, 315

\refer{}Hendry, P.~C., Lawson, N.~S., Lee, R.~A.~M.,
McClintock, P.~V.~E., \& Williams, C.~H.~D., in Formation and
Interactions of Topological Defects, ed.\ A.~C. Davis \& R.~N.
Brandenberger (Plenum, in press)

\refer{}Hindmarsh, M.~B., \& Kibble, T.~W.~B. 1995, Rep.\ Prog.\ Phys. 58, 477

\refer{}Hohenberg, P.~C., \& Halperin, B.~L. 1977, Rev.\ Mod.\ Phys.,
43, 435.

\refer{}Kibble, T.~W.~B. 1976, J.~Phys.\ A: Meth.\ Gen., 9, 1387

\refer{}Kibble, T.~W.~B. 1980, Phys.\ Rep., 67, 183.

\refer{}Kibble, T.~W.~B., \& Vilenkin, A. 1995, preprint
Imperial/TP/94-95/11 (TUTP 95-2)

\refer{}Leggett, A.~J. 1980, Progr.\ Theor.\ Phys.\ Suppl.,
69, 80

\refer{}Linde, A. 1990, Particle Physics and Inflationary Cosmology
(Harwood, New York)

\refer{}Linde, A.~D. 1982, Phys.\ Lett., 108B, 389.

\refer{}Lynton, E.~A. 1969, Superconductivity (Menuthen, London)

\refer{}Mineev, V.~P., Salomaa, M.~M., \& Lounasmaa, O.~V. 1987,
Nature, 326, 367

\refer{}Nielsen, H.~B., \& Olesen, P. 1973, Nucl.\ Phys., B61, 45

\refer{}Neumeier, J.~J., \& Zimmermann, H.~A., 1993, Phys. Rev. B47, 8385

\refer{}Neumeier, J.~J., 1994, Physica C233, 354

\refer{}Onsager, L. 1949, Nuovo Cim., 6, Suppl.~2, 249

\refer{}Rudaz, S., \& Srivastava, A.~M.  1993, Mod.\ Phys.\ Lett.,
A8, 1443

\refer{}Rudaz, S., Srivastava, A.~M., \& Varma, S. 1994, Probing the
Flux-Tube Network Produced in a Superconducting Phase Transition,
ITP preprint \#20.

\refer{}Ruutu, V.~M.~H., Eltsov, V.~B., Gill, A.~J., Kibble, T.~W.~B.,
Krusius, M., Makhlin, Y.~G., Placais, B., Volovik, G.~E., and Wen Xu,
Nature, submitted

\refer{}Salamon, M.~B., 1988, p. 39 in Physical Properties of High Temperature
Superconductors I, D.~M. Ginsberg, ed.

\refer{}Salomaa, M.~M. 1987, Nature, 326, 367

\refer{}Salomaa, M.~M.  in Formation and Interactions of
Topological Defects, ed.\ A.~C. Davis \& R.~N. Brandenberger
(Plenum, in press)

\refer{}Salomaa, M.~M., \& Volovik, G.~E.  1987,  Rev.\ Mod.\ Phys., 59,
533

\refer{}Tate, J., Cabrera, B., \& Felch, S.~B.  1984 in LT-17, ed.
U.~Eckern, A.~Schmid, W.~Weber, \& H. W\"uhl (Elsevier,
Amsterdam)

\refer{}Tilley, D.~R., \& Tilley, J. 1986, Superfluidity and
Superconductivity, second edition (Hilger, Boston)

\refer{}Tinkham, M. 1985, Introduction to Superconductivity
(Krieger, Malabar, Florida)

\refer{}Vachaspati, T. \& Vilenkin, A. 1984, Phys.\ Rev.\ D16, 1762 

\refer{}Vilenkin, A. 1981, Phys.\ Rev.\ D24, 2082.

\refer{}Vilenkin, A. 1985, Phys.\ Rep.\ 121, 264.

\refer{}Vilenkin, A., \& Shellard, E.~P.~S. 1994, Cosmic Strings and
Other Topological Defects (Cambridge University Press,
Cambridge)

\refer{}Vinen, W.~F. 1957, Proc.\ Roy.\ Soc.\ (London), A242,
433

\refer{}Vinen, W.~F.  1969, Chap.~20 in Superconductivity, Vol.~2,
ed.\ R.~D. Parks (M.~Dekker, New York)  (CHECK!!!)

\refer{} Volovik, G.~E. 1992, Exotic Properties of Superfluid He$^3$
(World Scientific, Singapore)

\refer{}Werthamer, N.~R. 1969, in Superconductivity, R.~D. Parks, ed.
(Dekker, New York)

\refer{}Wheeler, J.~A.,  \& Zurek, W.~H. 1983, eds., Quantum Theory
and Measurement (Princeton University Press, Princeton)

\refer{}Williams, G.~A. 1993, J.~Low.\ Temp.\ Phys., 33, 1079

\refer{}Yurke, B. 1995, in Formation and Interactions of Topological
Defects, ed.\ A.~C. Davis \& R.~N. Brandenberger (Plenum)

\refer{}Zeldovich, Ya.~B.,  Kobzarev, I.~Yu., \& Okun, L.~B. 1974, Zh.\
Eksp.\ Teor.\ Fiz., 67, 3; 1975, Sov.\ Phys.---JETP, 40, 1

\refer{}Zurek, W.~H. 1984, Experimental Cosmology: Strings in
Superfluid Helium, Los Alamos preprint LA-UR-84-3818

\refer{}Zurek, W.~H. 1985, Nature, 317, 505

\refer{}Zurek, W.~H. 1991, Physics Today, 44, 36

\refer{}Zurek, W.~H. 1993, Acta Physica Polonica B, 24, 1301

\refer{}Zurek, W.~H. 1994, Nature, 368, 292

\refer{}Zurek, W.~H. 1995, in Formation and Interactions of
Topological Defects, ed.\ A.~C. Davis \& R.~N. Brandenberger
(Plenum, New York)

\vfill\eject
\parindent 20pt
\parskip 11pt plus 2pt minus 1pt

Fig.~1.---Effective potential in Landau-Ginzburg model of second
order phase transition, $V(\varphi) = \alpha |\varphi|^2 + {\beta
\over 2}|\varphi|^4$.  Above the phase transition temperature $T_c$
(Fig.~1a) (coefficient $\alpha > 0$), there is just a single minimum for
$V(\varphi)$ at $\langle \varphi\rangle = 0$.  Below the phase
transition the minimum is degenerate and corresponds to the
expectation value of $\langle\varphi\rangle = \sigma =
\sqrt{|\alpha|/\beta}$, with $\Delta V = V (0) - V(\sigma) =
\alpha^2/2\beta$.

Fig.~2.---Phase diagram of He$^4$.  Liquid HeI is known as ``normal,''
while HeII is superfluid.

Fig.~3.---Structure of the superfluid vortex.  (a)~The density of the
superfluid increases outward from a ``normal'' core on a scale set
by the correlation length $\xi$.  Velocity of the superfluid falls off
with the distance so that at every radius the topological constraint
implied by Eqs.~(2.27) and~(2.36) is satisfied.  This leads to
one-dimensional vortex lines, shown in Fig.~3b.

Fig.~4.---Schematic trajectory of the pressure quench which can
induce a rapid phase transition from normal He$^4$ into the
superfluid ($a$).  As the $\lambda$-line is traversed, equilibrium
relaxation timescale $\tau$, Eq.~(2.45) diverges (6).  Thus, as the
quench proceeds at a finite pace, the order parameter will not be
able to adjust any more when the relaxation timescale becomes
comparable to the time when the critical temperature is reached.
This leads to a freezeout timescale $\hat t$; within the time interval
$\left[-\hat t, \hat t\right]$ the quench is in effect instantaneous.
Thus, the correlation length $\hat\xi$ corresponding to the instant
$\hat t$ will be frozen out and will set the density of topological
defects.

Fig.~5.---Formation of topological defects in the course of a rapid
phase transition.  Above the phase transition two-dimensional
order parameter fluctuates around the minimum of the potential,
assuming instantaneously values which ``point'' in approximately the
same direction on the ($\Psi_x, \Psi_y$) plane in the domains of
correlation-length size $\xi$.  If this configuration
was frozen out by an instantaneous phase transition, the symmetric
vacuum would be trapped, resulting in formation of topological
defects.

Fig.~6.---Actual quench trajectories in He$^4$ (after Hendry \etal,
1995).  Trajectory ($A$) never crosses or even approaches the
$\lambda$-line, and the corresponding quench does not lead to
vortex line production.  Trajectories which cross the $\lambda$-line
[such as ($C$)] result in copious vortex line densities.  Trajectories
which do not cross the $\lambda$-line, but come very close to it [as
is the case for ($B$)], also result in vortex line production, perhaps
as a result of a freezeout of thermal population of vortices which
may exist in the vicinity of the phase transition temperature.

Fig.~7.---Phase diagram of He$^3$ with a schematic indication of 
a few possible quench trajectories. In contrast to He$^4$, normal-superfluid
quench in Helium 3 involves increase of pressure. One could also contemplate
quenches from solid to superfluid for both isotopes pf Helium.

Fig.~8.---Rapid quench in an annulus.  Domain size $\hat\xi$ will now
determine the typical velocity of the superflow.  When $\hat\xi >
2r$, the problem is reduced to the evolution of a two-dimensional
(i.e., complex) order parameter in a one-dimensional space and is
illustrated in Fig.~9.

Fig.~9.---Evolution of the order parameter in the course of a quench
in a thin annulus can be illustrated with a version of the Argand diagram
shown above. Real and imaginary parts of the order parameter are plotted in the 
radial and in the vertical direction as a function of position along the 
circumference of the annulus. For a sufficiently thin annulus order parameter
will depend on only one variable---it will vary only with the location along 
the circumference of the annulus, which is shown as a greu ``doughnut''. 
An example of a possible instantaneous state is shown with the
black line. Above the phase transition temperature order
parameter $\Psi$ would fluctuate about the energetically favored $\Psi =
0$, changing its value significantly on a spatial scale given by the
correlation length $\xi$.  However, after the symmetry is broken typical
value of $|\Psi|$ will be set by the minimum of the potential at
$\sigma = \sqrt{|\alpha|/\beta}$, and the transitions through $\Psi =
0$ will become unlikely, as they require activation energy. Consequently, 
the initial configuration left by the quench will smooth out and be able 
to partialy unwind (oppositely oriented twists of the spiral will cancel). 
The leftover winding number (given by the nember of times the black line 
wraps around) will be stabilized and will result 
in a ``permanent'' superflow (in He$^4$) or supercurrent (in superconductors).

\bye